\documentclass[3p]{elsarticle}
\pdfoutput=1
\usepackage{epsfig}
\usepackage{amsmath}
\usepackage{amssymb}
\usepackage{wasysym}

\def\beq{\begin{equation}}
\def\eeq{\end{equation}}
\def\ba{\begin{eqnarray}}
\def\ea{\end{eqnarray}}
\def\bal{\begin{align}}
\def\eal{\end{align}}
\def\bxi{\boldsymbol{\xi}}
\def\bnab{\boldsymbol{\nabla}}
\def\calP{{\cal P}}

\begin{document}

\title{Non-radial Oscillations in Rotating Giant
  Planets with Solid Cores: Application to Saturn and its Rings}

\author[jf]{Jim Fuller\corref{cor1}}
\ead{derg@astro.cornell.edu}

\author[jf]{Dong Lai}

\author[jf]{Natalia I. Storch}

\cortext[cor1]{Corresponding author}

\address[jf]{Center for Space Research, Department of Astronomy, Cornell University, Ithaca, NY 14853, USA}

\label{firstpage}

\begin{abstract}

Recent observations have revealed evidence for the global acoustic
oscillations of Jupiter and Saturn. Such oscillations can potentially
provide a new window into the interior structure of giant planets.
Motivated by these observations, we study the non-radial oscillation
modes of giant planets containing a solid core. Our calculations
include the elastic response of the core and consider a wide range of
possible values of the core shear modulus. While the elasticity of the
core only slightly changes the frequencies of acoustic modes (including
the f-modes), which reside mostly in the fluid envelope, it adds two
new classes of shear modes that are largely confined to the core.  We also calculate the effects of the Coriolis force on the planetary oscillation modes. In addition to changing the mode frequencies, the Coriolis force can cause the shear modes to mix with the f-modes.  Such mixing occurs when the frequencies of the shear mode
and the f-mode are close to each other, and results in ``mixed modes''
that have similarly large surface displacements and gravitational
potential perturbations, but are slightly split in frequency. We
discuss our results in light of the recent work by Hedman \& Nicholson
(2013), which revealed the presence of density waves in Saturn's
C-ring that appear to be excited by the gravitational perturbations
associated with the f-mode oscillations within Saturn.  We find that
the fine splitting in wave frequencies observed in the rings can in
principle be explained by the rotation-induced mixing between core
shear modes and f-modes, possibly indicating the presence of a solid
core within Saturn. However, in our current calculations, which assume
rigid-body rotation and include only first-order rotational
effects, significant fine-tuning in the planetary model parameters is
needed in order to achieve these mode mixings and to explain the
observed fine frequency splitting. We briefly discuss other effects
that may modify the f-modes and facilitate mode mixing.

\end{abstract}


\maketitle

\section{Introduction}
\label{intro}

Despite enormous advances in precision measurements of various global
quantities of giant planets in our Solar System (mass, radius,
rotation rate, gravitational moments, oblateness, etc.), our knowledge
of the interior structure of these planets is rather limited.  One
uncertainty is the size of the central core, with estimates
in the range of $\sim (0-10)M_\oplus$ (Guillot 2005) and $\sim
(14-18)M_\oplus$ (Militzer et al.~2008) for Jupiter, and $\sim
(9-22)M_\oplus$ (Guillot 2005) for Saturn.  Another uncertainty
concerns the mixing and possible stratification of heavy elements in
the core and fluid envelope (Stevenson 1985; Leconte \& Chabrier 2012)

Global seismology is a promising technique for probing the internal
structures of stars and planets. Indeed, the internal structure of the
Earth, Moon, Sun, and numerous types of stars has been constrained
primarily via measurements of global oscillations (see Unno et
al.~1989 for stellar oscillations, Dahlen \& Tromp 1998, hereafter
DT98, for a comprehensive description of the techniques of Earth
seismology, and Chaplin \& Miglio 2013 a review of recent developments
in asteroseismology). Unfortunately, direct detection of global
oscillations in giant planets is extremely difficult because the
oscillations produce negligible luminosity perturbations and have
small surface displacements (radial surface displacements are likely
on the order centimeters). Recently, Gaulme et al.~(2011) reported the
detection of acoustic modes (p-modes) in the radial velocity data of
Jupiter, but the quality of the data was insufficient to provide new
constraints on Jupiter's interior model.

Saturn's ring system offers a unique opportunity to perform planetary
seismology, because even mild gravitational perturbations associated
with the planet's oscillation modes can generate density waves that
propagate through the rings. Marley \& Porco (1993) investigated this
idea in detail, arguing that some of the unexplained wave features (Rosen et al.~1991)
in Saturn's C and D-rings were produced at Lindblad resonances with the gravitational perturbations associated
with Saturn's oscillation modes. However, the existing {\it Voyager}
data was insufficient to measure the properties of the waves, and so
their seismic utility was limited.

Recently, Hedman \& Nicholson (2013) (hereafter HN13) used {\it Cassini} 
occultation data (see Colwell et al. 2009; Baillie et al.~2011)
to measure the radial location of Lindblad resonance ($r_L$),
azimuthal pattern number ($m$), and angular pattern frequency
($\Omega_p$) of several waves in Saturn's C-ring. We retabulate the
results of HN13 in Table 1. HN13 demonstrated that these waves can
not be excited by resonances with any of Saturn's (known or unknown)
satellites, but are compatible with being excited by low degree prograde sectoral
($l=|m|=2,3,4$) fundamental oscillation modes (f-modes) of Saturn (as
predicted by Marley 1991 and Marley \& Porco 1993). Throughout this paper, we adopt the convention that perturbations have the form $e^{im\phi+i\sigma t}$, with the inertial frame mode frequency $\sigma=|m|\Omega_p>0$.

Intriguingly, HN13 found what appeared to be a \textquotedblleft fine splitting" in
the mode frequencies: Instead of one $m=-2$ wave excited by Saturn's
$l=2$, $m=-2$ (prograde) f-mode,
there are {\it two} discrete waves with a frequency difference of
about $4\%$; instead of one $m=-3$ wave excited by the $l=3$, $m=-3$
f-mode, there are {\it three} waves with a frequency difference of
$(0.1-0.3)\%$.  No fine splitting was observed for the $m=-4$ wave,
and no waves with $m<-4$ were observed.

\begin{table*}
\begin{center}
\caption{Properties of the waves in Saturn's C-ring measured by
  HN13. The waves have the form $e^{im\phi+i\sigma t}$, with the wave
  frequency $\sigma=|m|\Omega_p$.  Resonant locations are measured
  from Saturn's center, and are taken from Baillie et al.~(2011). The
  value of $|\delta \tau|$ is the approximate maximum semi-amplitude
  of the optical depth variation associated with each wave.}
\begin{tabular}{@{}ccccc}
\hline\hline
Wave & Resonant location & $m$ & $\Omega_p$ (deg/day) & $|\delta \tau|$ \\
\hline
W80.98 & 80988 km & -4 & 1660.3  & $0.09$ \\
\hline
W82.00 & 82010 km & -3 & 1736.6  & $0.07$ \\
\hline
W82.06 & 82061 km & -3 & 1735.0  & $0.21$ \\
\hline
W82.21 & 82209 km & -3 & 1730.3  & $0.15$ \\
\hline
W84.64 & 84644 km & -2 & 1860.8  & $0.09$ \\
\hline
W87.19 & 87189 km & -2 & 1779.5  & $0.14$ \\
\hline\hline
\end{tabular}
\end{center}
\end{table*}

The origin of these fine splittings is puzzling. Saturn rotates rapidly (with a spin period of $10.6$~hours), which splits the f-mode of a given degree $l$ into multiples with azimuthal order
$m=-l,-l+1,...,l-1,l$.  For a given $m$, there are many f-modes that
correspond to different degrees $l$ in the norotating configuration,
but these modes differ in frequency by order unity (for small $l$),
much larger than the observed splitting. Rotation can also introduce
other Coriolis-force-supported modes (inertial modes and Rossby
modes), but these modes all have frequencies (in the rotating frame) less than $2\Omega_s \simeq 1630$ deg/day (where $\Omega_s$ is Saturn's rotation rate), which are smaller than
the f-mode frequencies. We discuss some other possible effects in Section \ref{discconc}.

In this paper, we explore the properties of oscillation modes of giant
planets that contain a solid core. We have two goals. (1) Previous
calculations of the oscillation modes of giant planets have been
restricted to pure fluid models, with or without a dense core (e.g.,
Vorontsov \& Zharkov 1981, Vorontsov 1981, Marley 1991, Wu 2005, Le
Bihan \& Burrows 2012). Although the global oscillations of the solid
Earth are well studied (DT98) and there have been some studies on the
effects of elasticity of the solid cores/crusts in white dwarfs and
neutron stars (e.g., Hansen \& Van Horn 1979, McDermott et al. 1988,
Montgomery \& Winget 1999), to our knowledge, no previous works have
investigated the elastic response of a solid core in giant
planets. The elasticity of a solid core adds entire new classes of
modes that have previously been ignored and can also modify the
properties of fluid modes (such as f-modes), which may have observable
signatures. (2) We examine the possibility of rotational mixing
between elastic core modes and envelope f-modes. While the influences of 
rotation on the mode frequencies are well studied in stars (Unno et al.~1989) and have been included in previous works on giant planetary oscillations (e.g., Vorontsov 1981; Marley 1991), the possibility 
of rotation-induced mode mixing has not been investigated.
We show that such mode mixing can in principle lead to
the appearance of multiple oscillation modes having very similar frequencies and 
characteristics. Nevertheless, as we show in this paper, significant fine tuning of the planetary model parameters
is needed to produce the observed fine splitting of the waves in
Saturn's rings.

Our paper is organized as follows. In Section \ref{planetarymodel}, we
generate simple giant planet models that will serve as the basis of
our oscillation mode calculations. Section \ref{elastic} describes the
characteristics of oscillations in non-rotating planets, while Section
\ref{rotationcoup} investigates mode mixing in rotating
planets. In Section \ref{gravpot}, we calculate the effects of
oscillation modes on Saturn's rings, and we compare our results to the
observations of HN13. In Section \ref{discconc}, we summarize our results and discuss
other effects (such as differential rotation and magnetic fields)
that may modify the oscillation modes and influence mode mixing.

\section{Planetary Model}
\label{planetarymodel}

Since our goal is to understand the effect of core elasticity on the
oscillations of giant planets and to explore the possibility of
rotation-induced mode mixing, we will not use sophisticated giant
planet models with ``realistic'' equation of state 
(e.g., Guillot 2005) in this paper. Instead,
we will consider simple planet models composed of a
one-component solid core surrounded by a neutrally stratified fluid
envelope charaterized by a $n=1$ ($\Gamma=2$) polytropic equation of state. 
These models allow us to capture the basic properties of
giant planets without getting bogged down in uncertain details (e.g.,
helium rain out, liquid-metallic hydrogen phase transitions, core size
and composition, etc.).

To generate our planet models, we first construct a polytropic model
of index $n=1$ (so that the pressure is related to density as $P
\propto \rho^2$). We then add a solid core by choosing a core radius,
$R_c$, a dimensionless density enhancement $D$, and constant shear
modulus $\mu$ (the shear modulus of the fluid envelope is zero). The
density of material in the core is calculated by multiplying the
density of material with $r<R_c$ in the original polytropic model by
$D$. We then normalize the density profile so that the total
mass/radius equal the mass/radius of Saturn. With this density
profile, we compute the gravitational acceleration via $g= G
M(r)/r^2$, where $M(r)=\int^r_0 4\pi r^2 \rho dr$. We then assume the
planet is neutrally stratified at all radii such that the
Brunt-Vaisala frequency $N^2=0$.  The pressure $P$ is obtained by
integrating the hydrostatic equilibrium equation $dP/dr=-\rho g$, and
the bulk modulus $K$ is given by
\beq
\label{K}
K = -\rho g \bigg(\frac{d\ln \rho}{dr}\bigg)^{-1}.
\eeq 
The bulk modulus is related to the pressure $P$ via $K = \Gamma_1 P$,
with $\Gamma_1 = d \ln P/d \ln \rho$.

For the purposes of calculating adiabatic acoustic-elastic pulsations
in non-rotating spherically symmetric planets, a planetary model is
completely described by three quantities as a function of radius: the
density $\rho$, adiabatic bulk modulus $K$, and the shear modulus
$\mu$ (see elastic oscillation equations in Section \ref{elastic}). 
Thus, our models have four free parameters: the density profile index\footnote{Note that $n$ is the polytropic index (as in $P\propto \rho^{1+1/n}$) of the planet without a core; when a core is added, the 
pressure and density in the fluid envelope no longer satisfies
the polytropic relation.} $n$, the radius of the solid core $R_c$ (or more precisely,
the ratio of $R_c$ to the planet radius $R$), the core-envelope density
jump $D$, and the (constant) core shear modulus $\mu$.
These are ideally suited for understanding the basic characteristics
of acoustic-elastic oscillations in giant planets. 
The $n=1$ polytropic density profile will generate
realistic estimates of frequencies of p-modes propagating in the fluid
envelope of the planet. The values of $R_c$, $D$, and $\mu$ affect the
spectrum of modes propagating in the solid core, therefore,
observations of these modes could provide strong constraints on core
properties.

\begin{figure}
\begin{center}
\includegraphics[scale=0.4]{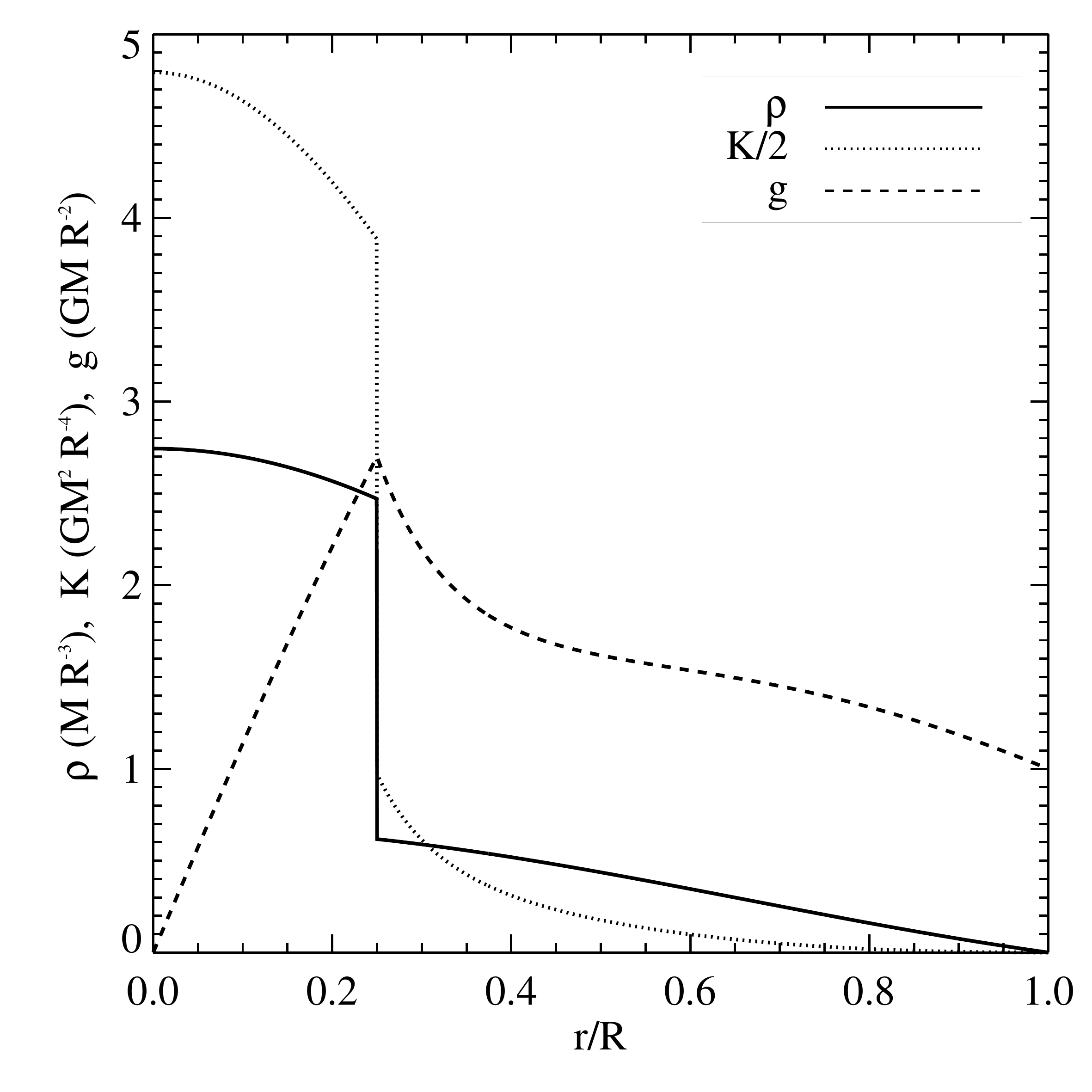}
\end{center} 
\caption{ \label{SatModel} Density, bulk modulus and gravity profiles
in our model of Saturn, with all quantities plotted in units with $G=M=R=1$. 
This model has an envelope density profile of a polytrope of index $n=1$, a core
radius $R_c=0.25R$ and a core density jump $D=4$.}
\end{figure}

Figure \ref{SatModel} displays the structure of our primary planet
model examined in this paper. It was constructed using $n=1$,
$R_c=0.25 R$, and $D=4$. It has a core mass $M_c=16 M_\oplus$, central
density $\rho_c = 7.1 \ {\rm g}/{\rm cm}^3$, and central pressure $P_c
= 2.2\times10^{12}$ Pa. These values are consistent with the
current observational constraints (see Guillot 2005).

\subsection{The Shear Modulus}
\label{shearmod}

As will be shown below, the value of the shear modulus in the solid
core determines the characteristics of elastic oscillation modes
within Saturn. The core of Saturn is likely composed of ices and
rocks. Assuming the core is at least partially solid, its shear
modulus may be determined by the properties of high pressure ices (of
which water ice is the dominant component).

Unfortunately, the shear modulus of water ice is unknown at the
pressures ($P \sim 10^{12}$Pa $=10^7$ bars) and temperatures ($T
\sim 8\times10^3$K, Guillot 2005) of Saturn's core. Asahara et
al. (2010) found $\mu \approx 8\times10^{10}$Pa for water ice at room
temperature and pressures of $6\times10^{10}$ Pa, well below the
central pressure of Saturn. The measurements of Asahara et al. (2010)
show the shear modulus of ice increasing toward larger pressures,
suggesting $\mu > 8\times10^{10}$Pa in the core of Saturn. However,
the shear modulus of many materials is also dependent on
temperature. In the solid core of the Earth, the shear modulus of iron
near its melting point is smaller than the shear modulus of iron at
lower temperatures and pressures (Laio et al. 2000).

Recent molecular dynamics simulations (Cavazzoni 1999, French 2009,
Militzer \& Wilson 2010, Militzer 2012, Hermann et al. 2011, Wang et
al. 2011) have predicted multiple new phases of ice at $P \gtrsim
10^{11}$Pa. In particular, ice is likely in either a superionic or
fluid phase at the core conditions of Saturn (French 2009, Wilson \&
Militzer 2012), depending on the core temperature. In the superionic
phase, the oxygen atoms form an ordered lattice, while the hydrogen
atoms diffuse freely through the lattice. It is possible that the
shear modulus of this superionic state is smaller than lower
temperature phases of ice in which the oxygen-hydrogen bonds
contribute to the strength of the lattice structure.

With such uncertainty in the shear modulus of material in a solid core
of Saturn, we take the shear modulus to be a free parameter. In this
work, we consider shear moduli in the range $2\times 10^8 {\rm Pa} <
\mu < 2\times 10^{12}$Pa, or $10^{-4}\lesssim \mu/P\lesssim 1$, which seems to be a reasonable
range given the above considerations. For simplicity, we also assume
the value of the shear modulus is constant throughout the core.

\section{Oscillations of Non-rotating Planets}
\label{elastic}

The presence of a solid core affects planetary oscillations through
the introduction of an elastic restoring force into the oscillation
equations. The elastic force is characterized by the shear modulus $\mu$,
whereas compressibility is characterized by the bulk modulus $K$.
In a homogeneous elastic medium, there are two types of waves:
(1) Sound waves (pressure waves or ``p-waves''): these are longitudinal waves,
with the displacement along the direction of wave vector ${\bf k}$, and 
the dispersion relation
\beq
\omega^2={K+4\mu/3\over\rho}k^2;
\eeq
(2) Shear waves (``s-waves''): these are transverse waves, with the displacement 
perpendicular to ${\bf k}$, and the dispersion relation
\beq
\omega^2={\mu\over\rho}k^2.
\label{eq:disper}
\eeq
The two polarizations of s-waves are designated SH (``shear horizontal'') and 
SV (``shear vertical'') in Earth seismology.
Obviously, shear waves do not exist in fluid regions.

In our planet models with a solid core, p-modes can propagate throughout
the planet, with different sound speeds in the core and in the liquid
envelope.  The introduction of elastic restoring forces allows for two
new types of modes to propagate in the solid core: spheroidal
shear modes and toroidal shear modes, for whom the elasticity is the
dominant restoring force.  The p-modes and spheroidal shear modes
(hereafter referred as ``s-modes'')\footnote{In the terminology of Earth 
seismology, the spheroidal shear modes correspond to SV modes
and the toroidal shear modes correspond to SH modes.}
are described by the displacement functions of the form
\beq
\bxi({\bf r}) = U(r) Y_{lm}(\theta,\phi) {\bf \hat{r}} + V(r)
r\bnab Y_{lm}(\theta,\phi).
\eeq
The toroidal shear modes (hereafter refereed as ``t-modes'') are described by
the toroidal displacement functions
\beq
\bxi({\bf r}) = W(r) \bnab \times \big[{\bf{r}} Y_{lm}(\theta,\phi)\big].
\eeq
The t-modes exhibit no radial displacement, no gravitational
perturbation, and are totally restricted to the solid regions of the
planet. The s-modes, on the other hand, are slightly affected by the pressure perturbation,
and can have a small, but finite gravitational perturbation.
Each mode oscillates at its eigenfrequency $\omega$ such that
\beq
\bxi({\bf r},t) = \bxi({\bf r})e^{i \omega t} \propto e^{i (m \phi + \omega t)}.
\eeq
With this convention, prograde modes with positive frequency have
$m<0$, while retrograde modes with positive frequency have $m>0$.

Introducing the elastic forces into the oscillation equations makes
them considerably more complicated. In  {\ref{app1}}, we list
the full adiabatic acoustic-elastic oscillation equations (see also
Dahlen \& Tromp 1998, Alterman et al. 1959, Hansen \& Van Horn 1979, Montgomery \& Winget
1999) and boundary conditions. The spheroidal oscillations are described by a system of
six linear coupled ordinary differential equations, while the toroidal
oscillations are decoupled from the spheroidal oscillation equations
and are described by two linear coupled ordinary differential
equations. We choose to normalize our mode eigenfunctions via their
inertia such that for any mode (indexed by $\alpha$),
\beq
\label{norm}
\int \!dV \rho \ \bxi_\alpha \cdot \bxi_\alpha^* = 1,
\eeq
with the integral extending over the volume of the planet.

\begin{figure}
\begin{center}
\includegraphics[scale=0.5]{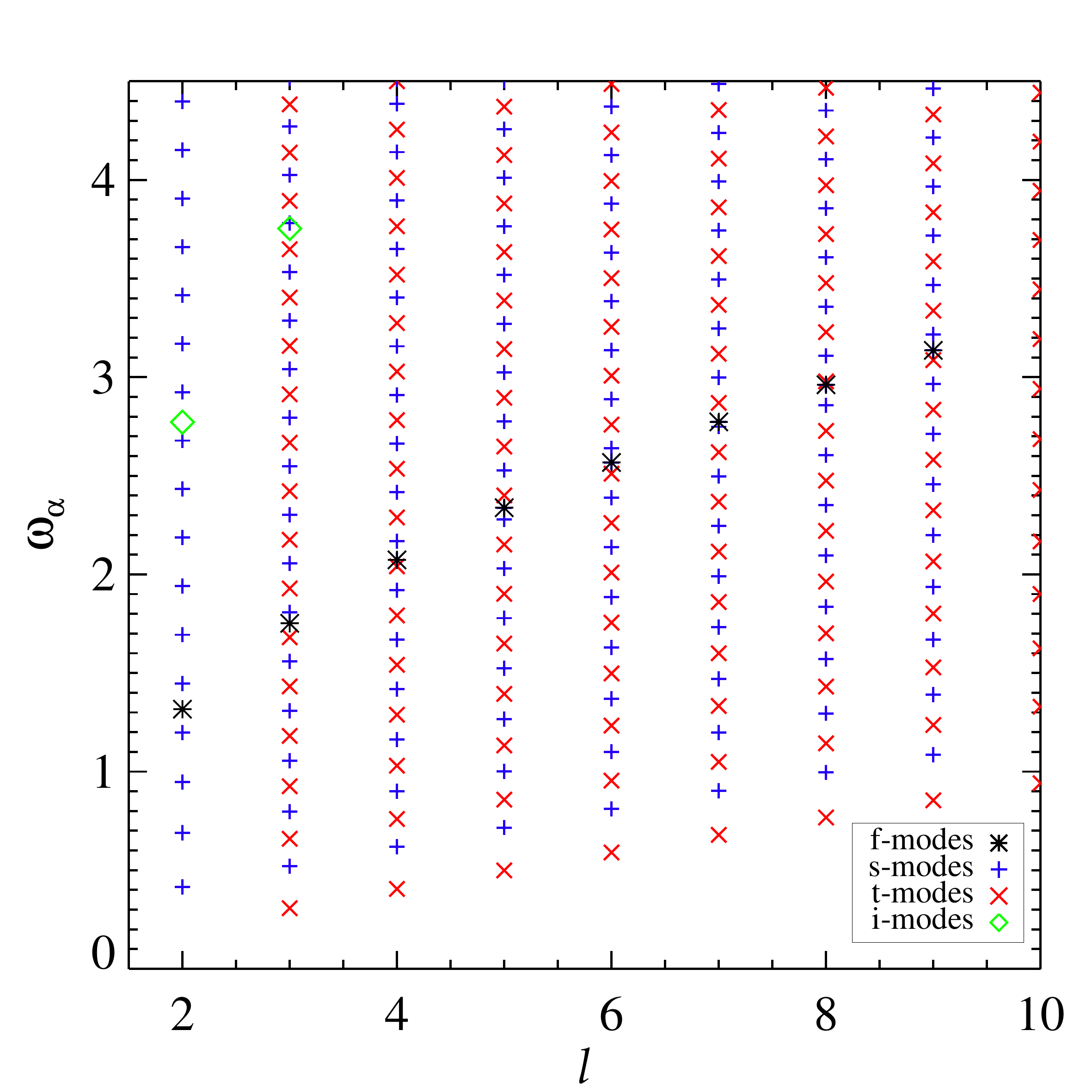}
\end{center} 
\caption{ \label{SatModes2} Oscillation mode spectrum of the planetary
  model with $n=1$, $R_c=0.25R$, $D=4$, and $\mu=1.6~{\rm GPa}$. The
  mode angular frequencies $\omega_\alpha$ (in units of
  $\sqrt{GM/R^3}$) are plotted as a function of the angular degree
  $l$. Only the fundamental p-modes (i.e. the f-modes) are shown, while higher-order p-modes are omitted for clarity. We have not included $l=2$ t-modes because they cannot mix with even parity modes
with $|m|>1$ (see Section \ref{modemix}). Note the nearly equal
frequency spacing for both s-modes and t-modes at all values of $l$.}
\end{figure}

\begin{figure}
\begin{center}
\includegraphics[scale=0.35]{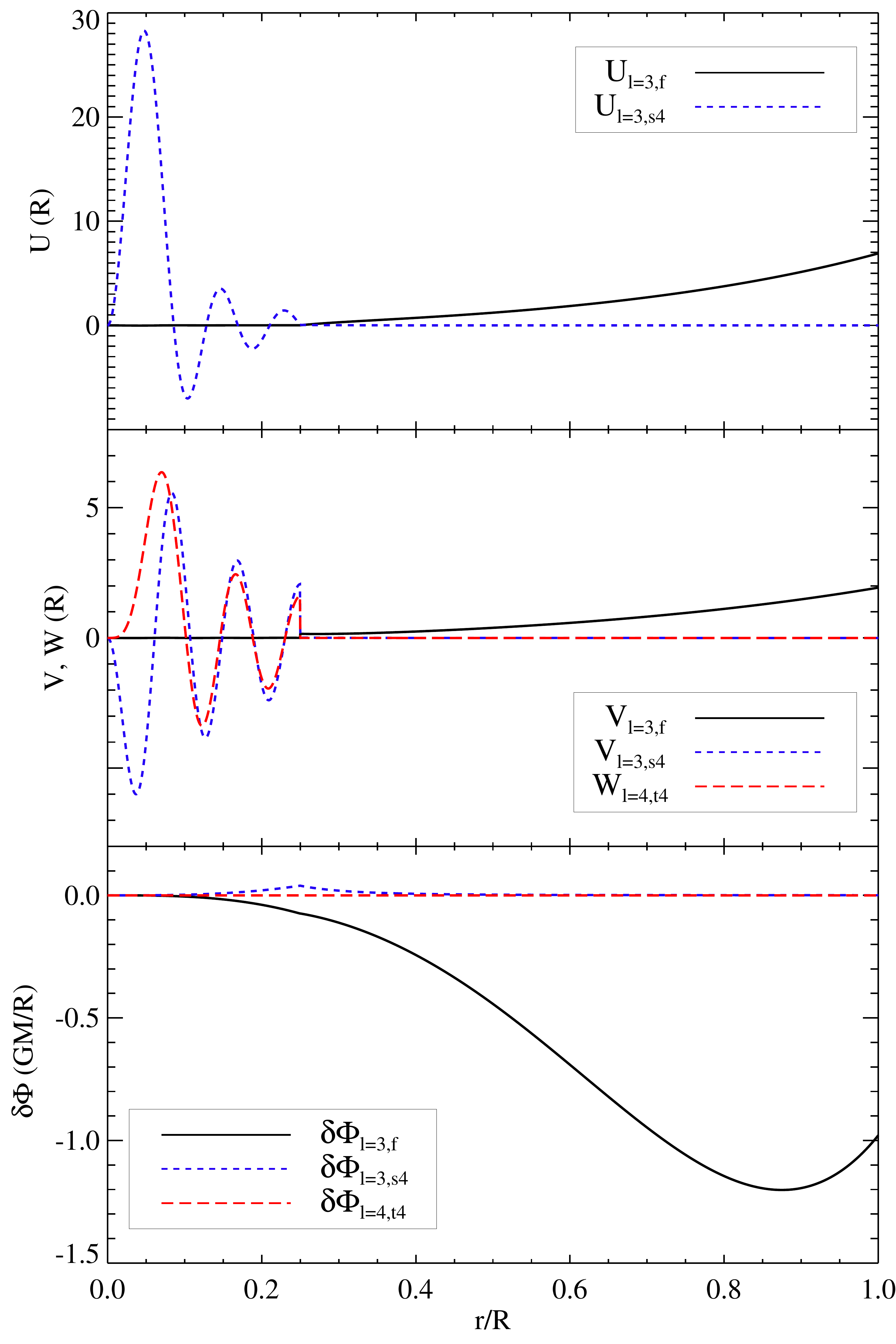}
\end{center} 
\caption{ \label{SatModeFunctions} Mode eigenfunctions in the
  planetary model with $n=1$, $R_c=0.25R$, $D=4$, and $\mu=1.6~{\rm
    Gpa}$. The radial displacement $U_\alpha(r)$ (top panel),
  horizontal displacement $V_\alpha (r)$ or $W_\alpha (r)$ (middle
  panel), and gravitational potential perturbation $\delta \Phi_\alpha
  (r)$ (bottom panel) are shown. The different lines correspond to the
  $l=3$ f-mode (black line, $\omega_\alpha=1.75$), a nearby $l=3$
  spheroidal mode ($s4$, blue dashed line, $\omega_\alpha=1.55$), and
  an $l=4$ toroidal mode ($t4$, red dashed line,
  $\omega_\alpha=1.54$).}
\end{figure}

\begin{figure}
\begin{center}
\includegraphics[scale=0.4]{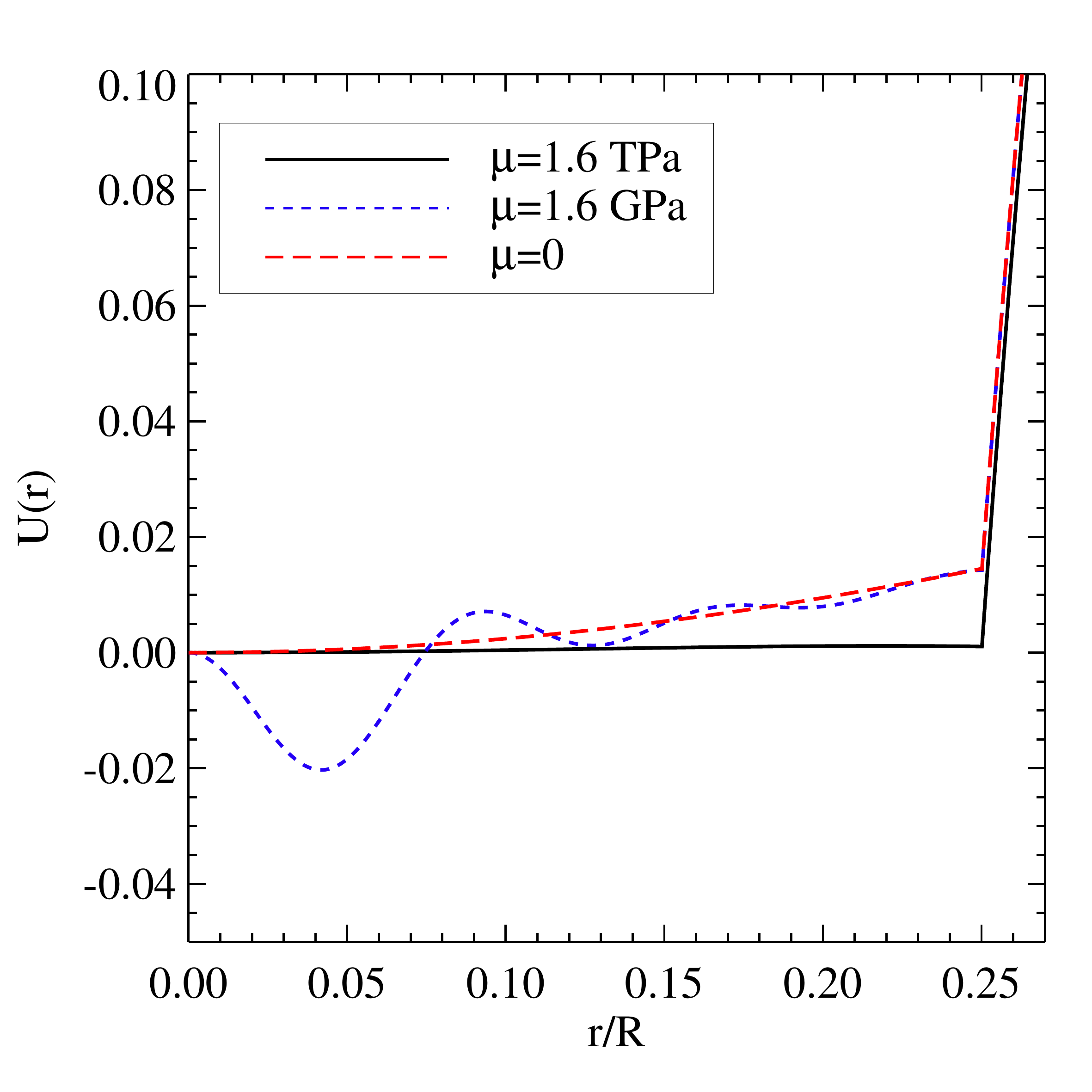}
\end{center} 
\caption{ \label{Fmodes} Radial displacements $U$ of the $l=3$ f-mode
  for planetary models with differing values of the core shear modulus
  $\mu$. The other model parameters are the same as in Figure \ref{SatModeFunctions}. The wave function for $\mu=1.6 \ {\rm GPa}$ obtains a wave-like character due to mixing with an s-mode (see Section \ref{modemix}). Note that the f-mode wave functions are nearly unaffected outside of the core.}
\end{figure}

\begin{figure}
\begin{center}
\includegraphics[scale=0.5]{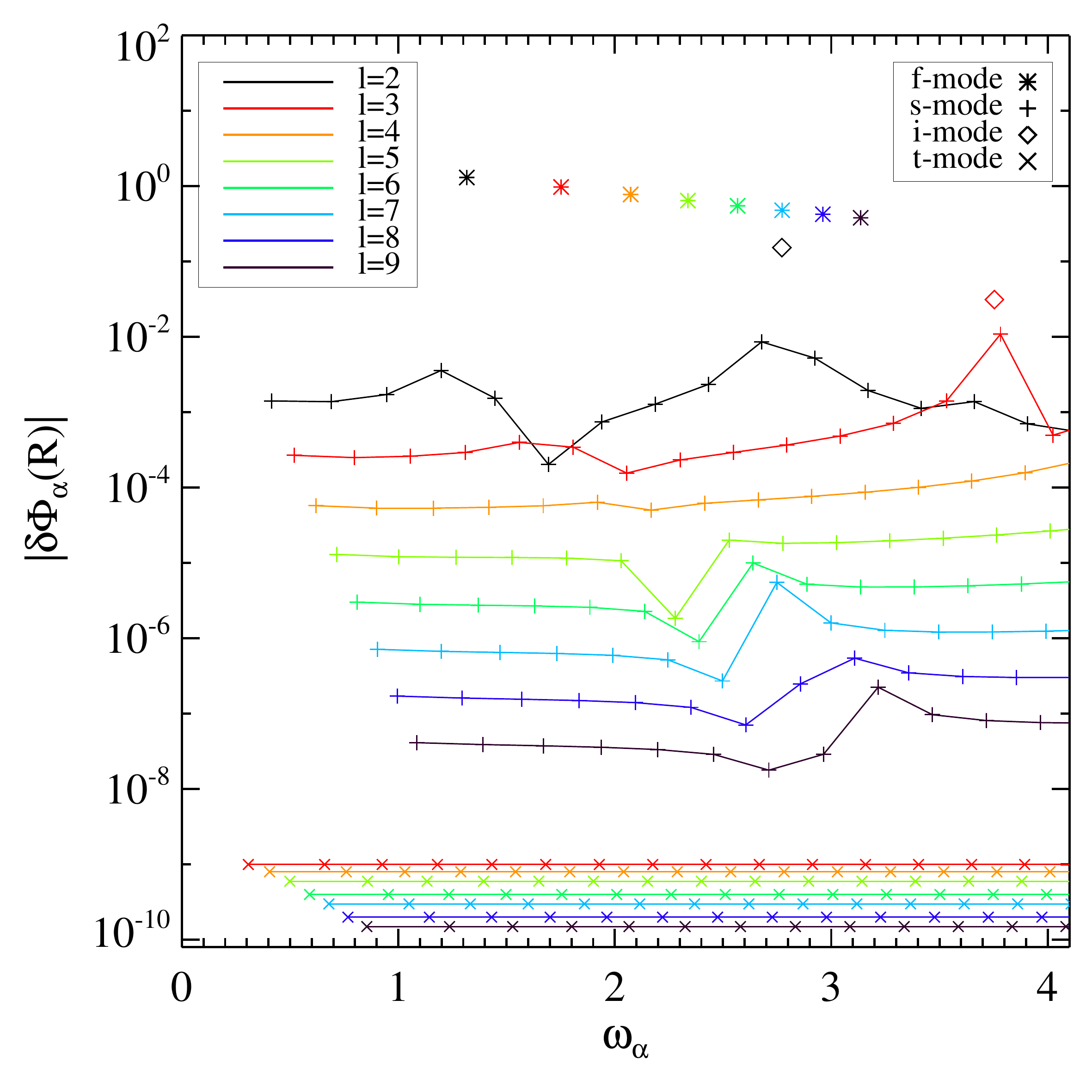}
\end{center} 
\caption{ \label{SatModes} Surface gravitational potential
  perturbation, $\delta \Phi_\alpha (R)$, as a function of mode
  frequency $\omega_\alpha$ in the planetary model with $n=1$,
  $R_c=0.25R$, $D=4$, and $\mu=1.6 \ {\rm GPa}$. S-modes and t-modes
  of equal value of $l$ have been connected by lines for clarity. The
  actual value of $\delta \Phi_\alpha (R)$ for the t-modes is exactly
  zero. The peaks in $\delta \Phi_\alpha (R)$ for s-modes are due to
  mixing with f-modes or i-modes. }
\end{figure}

Using relaxation techniques (see Press et al. 1998), we solve the acoustic-elastic oscillation equations for the spectrum of mode frequencies and eigenfunctions in a chosen planetary model.
Figure \ref{SatModes2} shows a plot of mode frequencies as a
function of $l$, while Figure \ref{SatModeFunctions} shows the
eigenfunctions of some oscillation modes of different types. 
Gas giant planetary models with a solid core and fluid envelope
support p-modes that are largely restricted to the fluid envelope, and
s-modes that are largely restricted to the solid core. The p-modes
typically have large surface displacements and gravitational potential
perturbations, while the s-modes have very small surface displacements
and potential perturbations. The t-modes are very similar to the core
s-modes, except that their displacement and potential perturbation are
exactly zero in the fluid envelope. 
\footnote{For a given $l$, the lowest-order s-mode corresponds
to the Rayleigh wave that travels near the solid surface, while
the lowest-order t-mode coresponds to the Love wave in the terminology of
terrestrial seismology.}

Both the s-modes and t-modes satisfy the local dispersion relation
(\ref{eq:disper}). The radial wave number is given by 
\beq
\label{disp}
k_r^2 = \frac{\rho}{\mu} \omega^2 - \frac{l(l+1)}{r^2}.
\eeq
These modes propagate at the shear speed $v_s = \sqrt{\mu/\rho}$. A core
with large density, small (but finite) shear modulus and large radius 
will support a dense spectrum of low frequency s-modes and t-modes. The frequency
spacing of these modes is
\beq
\Delta \omega \simeq \frac{\pi}{R_c} \sqrt{\frac{\mu}{\rho}}.
\eeq
This frequency spacing is nearly independent of the mode index
$\alpha$, and is dependent only upon the core radius, density, and
shear modulus. In the asymptotic regime, s-modes and t-modes of the
same $l$ are offset from one another by $\sim \Delta \omega/2$ because
of the differing boundary conditions at the core-envelope
boundary.\footnote{For modes completely confined to the solid core,
  the boundary conditions at the core-envelope boundary imply $U(r)
  \simeq 0$ for the s-modes but $dW(r)/dr \simeq 0$ for the
  t-modes. Therefore, the number of wavelengths differs by about $1/4$
  for s-modes and t-modes, accounting for the offset of $\sim\Delta
  \omega/2$ between s-modes and t-modes in Figure \ref{SatModes2}.}
Consequently, for s-modes of angular degree $l$, there exist t-modes
of $l\pm1$ with similar frequencies, and vice versa for t-modes.

The effect of a solid core on the properties of the f-mode and low
order p-modes is miniscule. The main reason is that f-modes and low
order p-modes have almost all of their inertia in the fluid
envelope. Consequently, the value of the shear modulus has essentially
no impact on the f-mode frequency (the the f-modes shown in Figure
\ref{Fmodes} differ by less than one part in $10^{5}$ in frequency),
radial surface displacement, or potential perturbation.\footnote{An
  exception to this rule is at frequencies very near the avoided crossings
  with s-modes or i-modes, see Section \ref{modemix}.} Nonetheless,
the value of the shear modulus does affect the f-mode wave function in
the core of the planet, as shown in Figure \ref{Fmodes}. For very
rigid cores (large $\mu$), incoming waves are reflected, and the
f-mode is excluded from the core. For softer cores (small but finite
$\mu$), the f-mode may obtain a wave-like structure inside the
core. Finally, for fluid cores, the f-mode penetrates into the core,
but with a small amplitude, due to the jump in density.

In the absence of rotation, the external gravitational perturbation will
be produced almost exclusively by envelope f-modes and low order
p-modes (see Figure \ref{SatModes}).\footnote{Moreover, only
  relatively low-degree (low value of $l$) modes will produce
  significant external gravitational perturbations because the
  strength of the perturbation outside the planet falls off as $\delta
  \Phi(r) = \delta \Phi(R) (R/r)^{l+1}$, where $r$ is the distance
  from the center of the planet.} 
The incompressive nature of low-order s-modes and t-modes creates 
zero (for t-modes) or very small density and gravity perturbations. Furthermore, because s-modes
and t-modes are largely confined to the core, they produce very small
fluid displacements at the surface. Thus, in the absence of rotation,
s-modes and t-modes should be nearly impossible to detect.

Because our planetary models are neutrally stratified, they do not
support g-modes. However, the discontinuity in density at the
core-envelope boundary supports a single interface mode for
each value of $l$. The frequency of this i-mode is given by
\beq
\omega_{i}^2 \approx \frac{\sqrt{l(l+1)}g_c}{R_c} \frac{\rho_b - \rho_a}{\rho_b + \rho_a},
\eeq
where $a$ and $b$ indicate that the quantity should be evaluated above and
below the interface, respectively, and $g_c=g(R_c)$. For density jumps
of $D \lesssim 2$ at the core-envelope boundary, the frequency of the
interface mode may be comparable to that of the 
f-mode.\footnote{Additional interface modes may also exist due to the
  presence of other density discontinuities in the planet. This
  may occur at the molecular-metallic hydrogen phase transition, the
  metallic hydrogen-molecular helium composition gradient, or at
  discontinuities in a differentiated core.} The surface displacements
and gravitational potential perturbations of the i-modes are typically
greater than s-modes but less than f-modes for our planetary models.

\subsection{Mode Mixing}
\label{modemix}

The distinction between different types of modes (s-modes, t-modes,
i-modes, and f-modes or p-modes) is not always clear. For a given
planetary model, there may exist modes of nearly equal $\omega_\alpha$
and identical $l$ that will mix and obtain characteristics of two
types of modes. This phenomenon is well known in Earth seismology
(DT98), and is frequently observed as mixed modes in red giant stars
(Chaplin \& Miglio 2013).

\begin{figure}
\begin{center}
\includegraphics[scale=0.45]{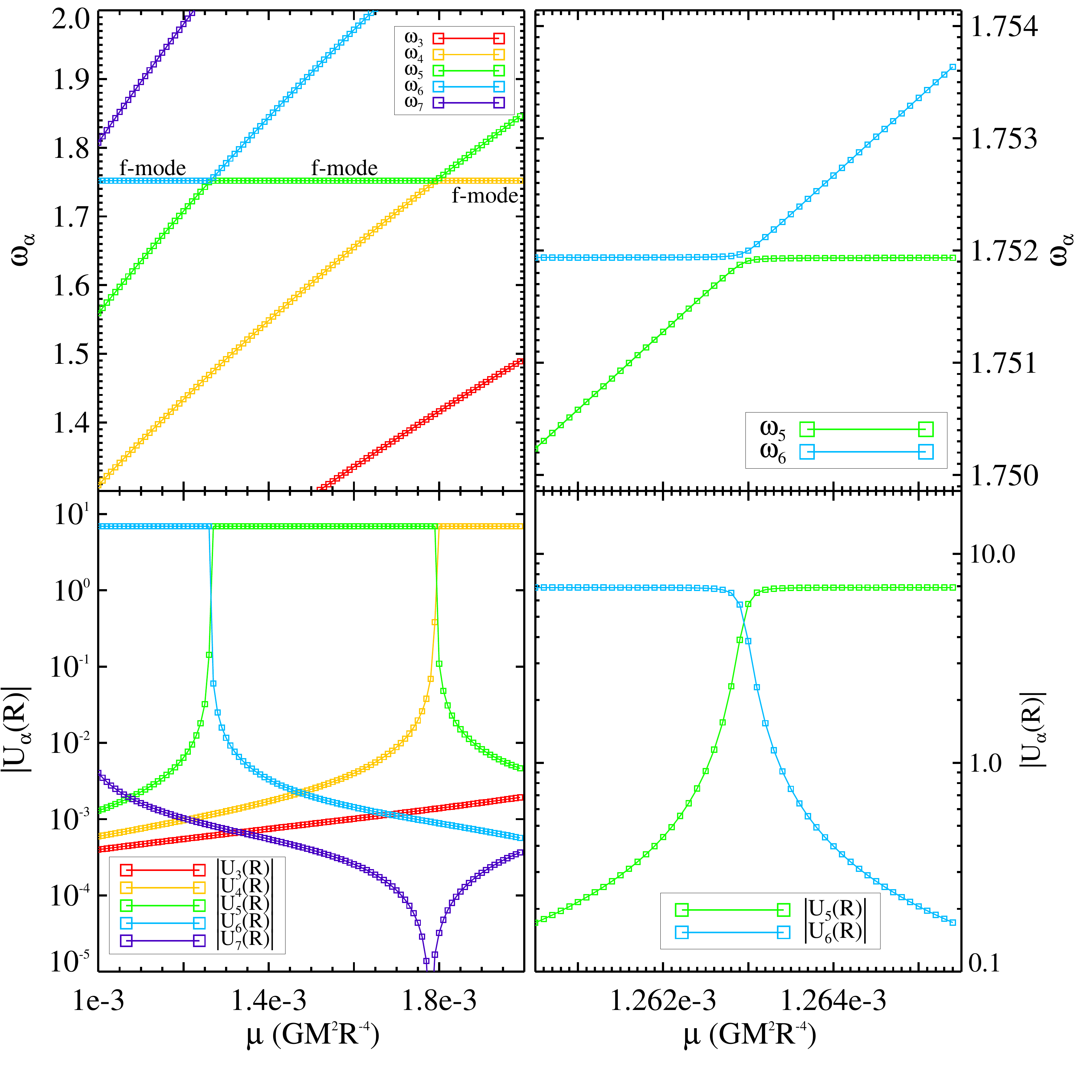}
\end{center} 
\caption{ \label{MuScan} The frequencies $\omega_\alpha$ (top two
  panels) and radial surface displacements $U_\alpha(R)$ (bottom two
  panels) of a few selected modes for our planet models (as depicted in Figure
  \ref{SatModel}), as a function of the value of the shear modulus
  $\mu$. Away from avoided crossings, the f-mode corresponds to the
  mode lying along the nearly horizontal line at $\omega_\alpha \simeq
  1.75$, while the other modes are core s-modes. The right two panels
  present a zoom-n view of an avoided crossing between an s-mode
  and the f-mode.}
\end{figure}

To understand the process of mode mixing in our non-rotating giant
planet models (we examine rotation-induced mode mixing in Section
\ref{rotationcoup}), we calculate the modes
for different values the shear modulus $\mu$.  The frequencies
$\omega_\alpha$ and surface displacements $U_\alpha(R)$ of some
selected modes are shown in Figure \ref{MuScan}. At certain shear
moduli $\mu=\mu_c$, the f-mode and an s-mode have nearly identical
frequencies. As the value of $\mu$ approaches $\mu_c$, the modes begin
to exchange character with one another, causing the f-mode to
penetrate into the core and the s-mode to penetrate into the fluid
envelope. At $\mu=\mu_c$, the modes reach a minimum frequency
separation and are equal superpositions of one another. As $\mu$
increases away from $\mu_c$, the mode frequencies diverge from one
another (note the frequencies are never exactly equal, resulting in an
\textquotedblleft avoided" crossing), having smoothly exchanged mode
character. If a planetary model happens to contain modes near these
avoided crossings, core s-modes may obtain substantially larger
surface displacements and gravitational potential perturbations.

However, we find that it is unlikely for an s-mode and an f-mode to be
near an avoided crossing in our planetary models, as evidenced by the
extremely narrow width of the avoided crossings in Figure
\ref{MuScan}. Avoided crossings between s-modes and i-modes have
larger frequency widths, but will be difficult to observe due to the
smaller surface displacements and potential perturbations of i-modes
relative to f-modes. Finally, avoided crossings between i-modes and
f-modes or p-modes have substantial frequency widths, but are unlikely
to be observed because of the small number of i-modes (our models have
only one i-mode for each value of $l$). We conclude that it is
unlikely to see mode mixing phenomena (in the absence of rotation) in
giant planets.

\section{Rotational Mode Mixing}
\label{rotationcoup}

Thus far, we have considered the adiabatic acoustic-elastic
oscillations of spherically symmetric, non-rotating planet models. However,
the giant planets in our solar system spin rapidly, and the effects of
rotation are quite important. For reference, Saturn's rotation period
of $\simeq 11~{\rm hr}$ corresponds to $\Omega_s \simeq 0.36
\ \Omega_{\rm dyn}$ (where $\Omega_{\rm dyn} = \sqrt{GM/R^3}$ is the
dynamical frequency of the planet), while the frequencies of the
f-modes are typically 
$\omega_f \sim \sqrt{l} \Omega_{\rm dyn}$. 
Previous studies (e.g., Marley 1991 and Vorontsov \& Zharkov
1981) have shown that including rotational effects is essential in
predicting the frequencies of f-modes of giant planets. Most 
studies of modes in rotating stars or planets utilize perturbation theory to obtain
corrections to the frequencies and eigenfunctions of the modes,
computed in powers of the small parameter $\lambda =
\Omega_s/\omega_0$, where $\omega_0$ is the unperturbed mode
frequency. To first order in $\lambda$, the only correction is
from the Coriolis force, which splits the $2l+1$ degenerate
eigenfrequencies (corresponding to the $2l+1$ values of $m$ for a
given $l$) of the unperturbed planetary model. The second order rotational
effects include the centrifugal force and the rotationally-induced
planetary oblateness. In the absence of a solid core, the envelope
p-modes are well separated in frequency, and non-degenerate
perturbation methods suffice when calculating the influence of
rotation.

However, in planetary models with a solid core, the addition of
s-modes and t-modes may cause the spectrum of modes to become dense
near the f-modes. In this case, the frequency spacing between modes
may be smaller than the rotational corrections, and non-degenerate
perturbation methods fail. Rotation not only shifts the frequencies of
oscillation modes, but can also induce strong mode mixing. Spheroidal
modes may acquire toroidal components, and vice versa. Both s-modes
and t-modes may mix strongly with the f-mode, and a more precise
treatment of the rotational effect is necessary.  In this paper, we
consider only the first order rotational effect (i.e., the Coriolis
force), and assume uniform planetary rotation.

For two modes (labeled $\alpha$ and $\alpha'$) to be mixed by the
Coriolis force, the mixing matrix element (see \ref{rotcoup})
\beq
\label{eq:Cdef}
C_{\alpha\alpha'} = i \int\!dV \rho \bxi_{\alpha}^* \cdot 
\big(\boldsymbol{\Omega_s}\times \bxi_{\alpha'}\big),
\eeq
where $\boldsymbol{\Omega_s}$ is the spin vector, must be nonzero.
This leads to some basic selection rules. First, only modes with $m=m'$ will mix. Second,
spheroidal modes only couple to other spheroidal modes with $l=l'$,
and likewise for toroidal modes. Finally, spheroidal and toroidal
modes couple to one another only if $l=l'\pm1$. Thus, 
although a spheroidal mode of angular degree $l$ does not couple directly to
other spheroidal modes of $l'=l\pm2$,\footnote{Second order rotational
effects introduce direct coupling between modes with $l=l'\pm2$, but we
consider only Coriolis coupling in this work.} they can couple
indirectly through intermediary toroidal modes. 
We do not consider rotational mixing with inertial or Rossby modes
(if they exist in Saturn's interior) because these modes have a
maximum frequency of $\omega=2\Omega_s$ in the rotating frame, and
thus have smaller frequencies than any of Saturn's f-modes. Our method
for calculating the effect of rotational mode mixing is outlined in \ref{perturb}. Here we describe only the basic ideas.

Consider a simple two mode system. In the absence of rotation, the two
eigenmodes have frequencies $\omega_1$ and $\omega_2$, with
eigenvectors ${\bf Z}_1$ and ${\bf Z}_2$. Including the Coriolis
correction, we project the rotationally modified eigenmodes onto the
original eigenmodes such that ${\bf Z} = a_1{\bf Z}_1 + a_2{\bf Z}_2$
(see \ref{perturb}). The eigensystem (equation \ref{mat11})
describing the mixed modes is
\beq
\label{mix1}
\begin{bmatrix} \bar{\omega}_1-\omega & C_{12} \\  C_{12}^* & \bar{\omega}_2-\omega \end{bmatrix} \left[ \begin{array}{c} b_1 \\ b_2 \end{array} \right] = 0 
\eeq
where $b_1 = \omega_1 a_1$, $b_2 = \omega_2 a_2$, and
$\bar{\omega}_1 = \omega_1 + C_{11}$ and
$\bar{\omega}_2 = \omega_2 + C_{22}$
are the usual rotation-corrected mode frequencies when mixing is neglected.
The Coriolis coupling coefficients $C_{\alpha\alpha'}$ are defined in equation 
\ref{eq:Cdef}. 
Defining $\Delta_{12} = \bar{\omega}_1 -\bar{\omega}_2$, equation 
(\ref{mix1}) can be written as
\beq
\label{mix2}
\Bigg(\begin{bmatrix} \bar{\omega}_1+\bar{\omega}_2 - 2 \omega & 0 \\ 0 & \bar{\omega}_1+\bar{\omega}_2 - 2 \omega \end{bmatrix} +  \begin{bmatrix} \Delta_{12} & 2C_{12} \\  2C_{12}^* & -\Delta_{12} \end{bmatrix} \Bigg) \left[ \begin{array}{c} b_1 \\ b_2 \end{array} \right] = 0.
\eeq
This clearly shows that strong mode mixing occurs only if
$2|C_{12}|\gtrsim |\Delta_{12}|$. The eigenfrequencies of 
equation (\ref{mix1}) or (\ref{mix2}) are
\beq 
\omega = \frac{\bar{\omega}_1 + \bar{\omega}_2}{2} \pm
\frac{1}{2} \sqrt{\Delta_{12}^2 + 4|C_{12}|^2}.
\eeq
The corresponding eigenvectors are
\beq
\label{bvec1}
{\bf b}_+ = \left[ \begin{array}{c} \cos \theta_{12} \\ x_{12}^*\sin \theta_{12} \end{array} \right]
\qquad {\rm and}\qquad
{\bf b}_- = \left[ \begin{array}{c} \sin \theta_{12} \\  -x_{12}^*\cos \theta_{12} \end{array} \right], 
\eeq
where $x_{12}\equiv C_{12}/|C_{12}|$,\footnote{The value of $C_{12}$ is real
for spheroidal-spheroidal and toroidal-toroidal mode coupling (i.e.,
  $x_{12}=\pm 1$ in this case), and is imaginary for
  spheroidal-toroidal mode coupling (i.e., $x_{12}=\pm i$ in this
  case).}
and we have defined the mode mixing angle $\theta_{12}$ via
\beq
\tan 2\theta_{12} = \bigg| \frac{2C_{12}}{\Delta_{12}} \bigg|.
\eeq

When $2|C_{12}| \ll |\Delta_{12}|$ or $\theta_{12}\ll 1$, mode mixing
is negligible, and the mode frequencies are simply $\bar\omega_1$ and
$\bar\omega_2$. With decreasing $|\Delta_{12}|$ or increasing $|C_{12}|$,
the mixing angle $\theta_{12}$ increases
above zero, approaching $\pi/4$ in the limit $2|C_{12}|\gg |\Delta_{12}|$.
In this limit, the modified mode eigenfunctions are equal
superpositions of the original mode eigenfunctions. Hence, if a core
mode is strongly mixed to the f-mode, it will obtain the f-mode
characteristics, including a much larger radial surface displacement
$|\xi_\alpha(R)|$ and gravitational potential perturbation $|\delta
\Phi_\alpha(R)|$.  A ``nearby'' mode not strongly mixed or 
degenerate with the f-mode may still have an enhanced gravitational 
potential perturbation, with
\beq
\delta \Phi_\alpha \simeq \frac{C_{{\rm f}\alpha}}{\Delta_{{\rm
      f}\alpha}} \delta \Phi_{\rm f}, 
\eeq
where $C_{{\rm f} \alpha}$ and $\Delta_{{\rm f}\alpha}$ are the
mixing strength and frequency separation between the f-mode and the
nearby mode. For planetary models in which the mode spectrum is dense near the
f-mode, we should expect to see a peak in $\delta \Phi_\alpha$
centered on the f-mode, with a frequency width at its peak of order 
$2|C_{{\rm f} \alpha}|$.

\subsection{Strength of Coriolis Mode Mixing}

The importance of mode mixing due to the Coriolis force is determined
by the value of $C_{\alpha\alpha'}$. For most modes, the value of
$|C_{\alpha\alpha'}|$ is largest when $\alpha=\alpha'$, i.e., due to
self-coupling. 
This self-coupling term is identical to the standard
rotationally induced frequency shift, i.e., if $\omega_\alpha$ is the
frequency of mode $\alpha$ without rotation, then the Coriolis effect
changes the frequency (in the rotating frame) to
\beq
\bar\omega_\alpha=\omega_\alpha+C_{\alpha\alpha}.
\eeq
For s-modes and t-modes, the self-coupling
coefficient has a value of $C_{\alpha\alpha} \approx m
\Omega_s/[l(l+1)]$, while its value is typically of order $C_{{\rm f}
  {\rm f}} \approx m \Omega_s/l$ for the f-modes in our planetary models. 
Thus, low-degree prograde sectoral ($m=-l$) modes near the f-mode are typically
reduced in frequency by a few $\times 10\%$ in Saturn due to the
effect of self-coupling.

In contrast to self-coupling, the mixing between core modes and envelope
modes is typically quite weak in our planetary models. The main reason
is simply that s-modes and t-modes have nearly all their inertia in
the core, while f-modes and p-modes have nearly all their inertia in
the envelope, and so the mode eigenfunctions do not have a large
overlap in any region of the planet. We find typical values
$|C_{\alpha\alpha'}| \lesssim 10^{-3} |m| \Omega_s$ for mixing between
core modes and f-modes of similar frequencies in our planetary
models. These values decrease with decreasing shear modulus $\mu$
because of the increasingly oscillatory wave functions of the s-modes
and t-modes. The value of $|C_{\alpha\alpha'}|$ also decreases with
increasing $l$ because f-modes of larger angular degree are confined
closer to the planetary surface and have less inertia in the
core. Therefore, the detuning $\Delta_{12}$ between a core mode and an
f-mode must be very small for appreciable mixing to occur.

We also note that the mixing between s-modes and t-modes can be
strong, with the values of $|C_{\alpha\alpha'}|$ rising as large as
the typical values for self-coupling. These large values of
$|C_{\alpha\alpha'}|$ occur for s-modes of angular degree $l$ and
t-modes of angular degree $l\pm1$ with the same number of nodes in
their radial eigenfunctions. In non-rotating models, these modes typically have $\omega_\alpha \simeq \omega_{\alpha'}$,
however, their frequencies must be split by at least $\sim 2|C_{\alpha\alpha'}|$ in rotating planets.
Such strong mixing implies that each low-frequency core mode is a strong
superposition of both spheroidal and toroidal components in a rapidly
rotating planet.

\subsection{Two and Three-Mode Rotational Mixing}
\label{2and3mode}

\begin{figure}
\begin{center}
\includegraphics[scale=0.35]{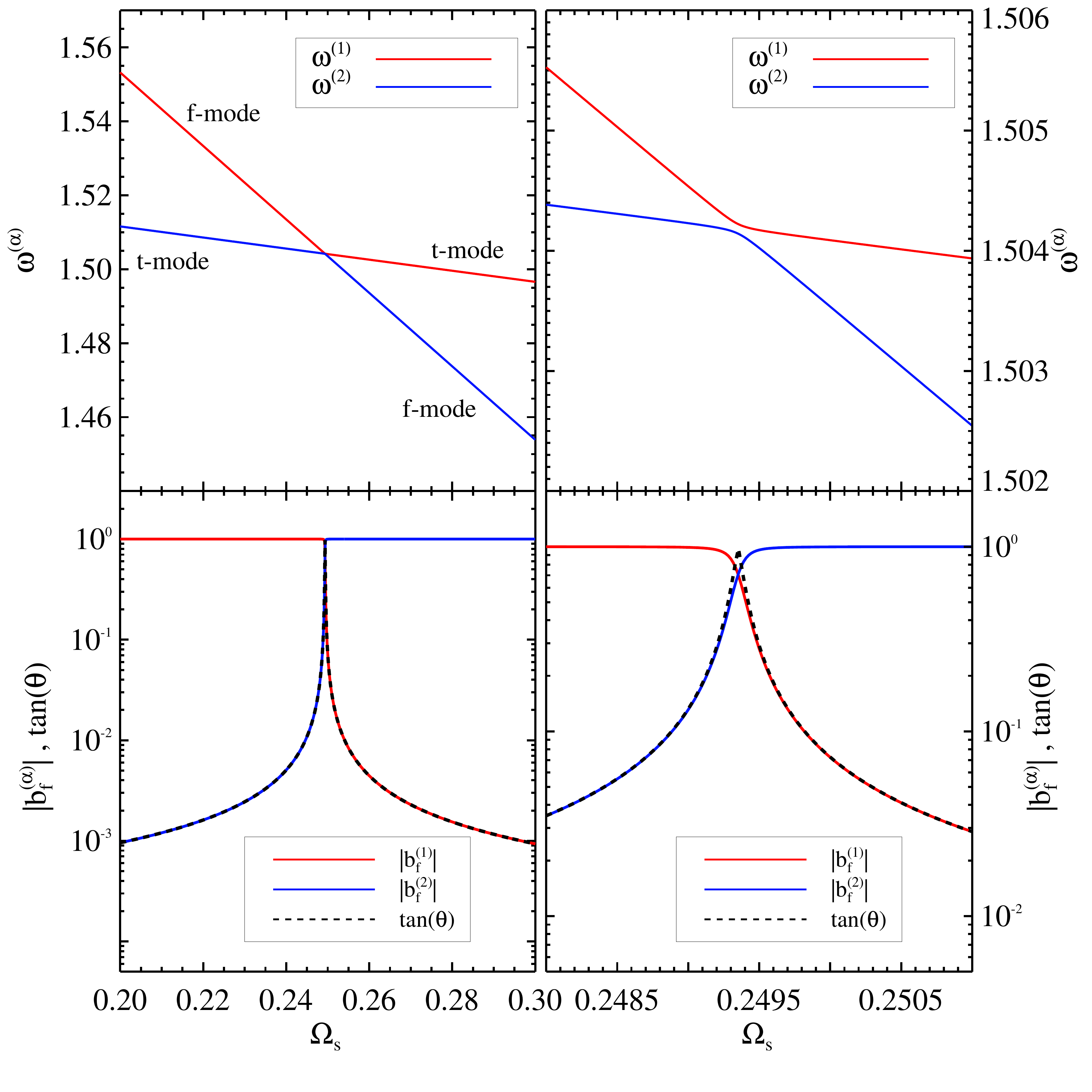}
\end{center} 
\caption{ \label{2mode} Top: The frequencies $\omega^{(\alpha)}$ of
  the rotationally mixed $l=3$ f-mode and an $l=4$ t-mode
  as a function of Saturn's spin
  frequency $\Omega_s$ (in units of $\sqrt{GM/R^3}$). Bottom: The
  projections $b^{(\alpha)}_{\rm f}$ of each rotationally mixed mode
  onto the f-mode. We have also plotted the mode mixing angle
  $\tan(\theta)$. The modes have an avoided crossing at $\Omega_s
  \simeq 0.25$, where they are maximally mixed. The \textquotedblleft
  f"-mode corresponds to $\omega^{(1)}$ for $\Omega_s \lesssim 0.25$, and
  corresponds to $\omega^{(2)}$ for $\Omega_s \gtrsim 0.25$. 
  The right panels show a zoom-in on the left panels.
  The eigenmode amplitudes have been normalized via $\sum_\beta |b^{(\alpha)}_\beta|^2 = 1$.}
\end{figure}

Figure \ref{2mode} shows an example of rotation-induced mixing between
two modes.
To make this figure, we consider the mixing between the $l=3$, $m=-3$ f-mode 
and the $l=4$, $m=-3$ t-mode depicted in Figure \ref{SatModeFunctions}. At most spin frequencies,
the modes are well-separated, and the mixing angle is small because of
the small value of $C_{{\rm f},\alpha}$. However, near a particular
spin frequency the modes become nearly degenerate and mix strongly
with one another. At maximum mixing, the modes consist of nearly equal
superpositions of the original f-mode and t-mode, and 
there is the ``avoided'' crossing behavior of the frequencies.
This is analogous to the mixing described in Section \ref{modemix}.
Note that for a given f-mode and a nearby mode, strong mixing occurs 
over a very narrow range of spin frequency, with the width $\Delta\Omega_s$
of order $|C_{{\rm f}\alpha}|$, and $|C_{{\rm f}\alpha}|/\Omega_s \ll 1$.

\begin{figure}
\begin{center}
\includegraphics[scale=0.4]{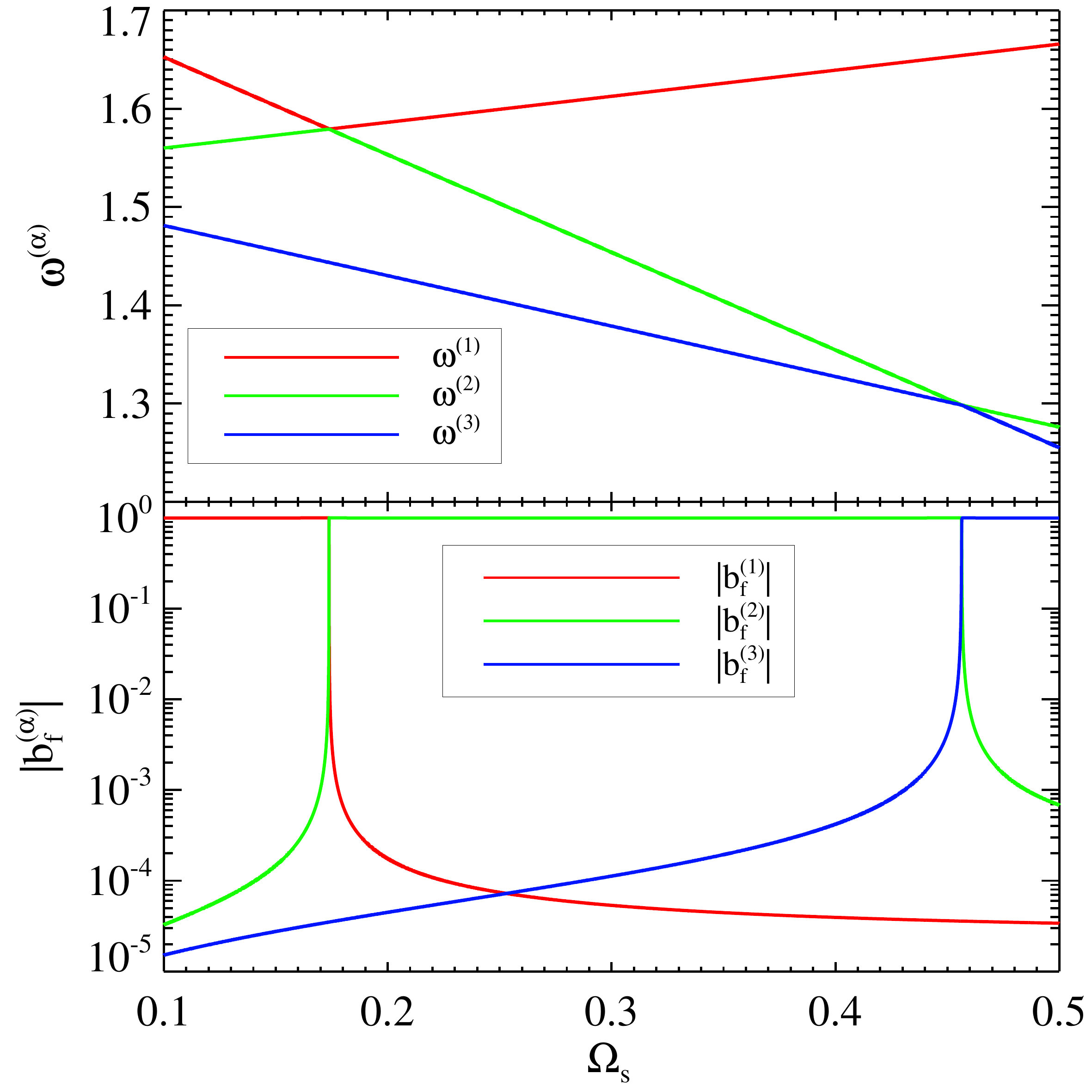}
\end{center} 
\caption{ \label{3mode} Same as Figure \ref{2mode}, but for a system
  of three mixed modes. Away from the avoided crossings, the
  \textquotedblleft f"-mode corresponds to the line with the steepest
  slope. The \textquotedblleft s"-mode and \textquotedblleft t"-mode
  are strongly mixed at all values of $\Omega_s$, but only mix
  significantly with the \textquotedblleft f"-mode near the avoided
  crossings.}
\end{figure}

Three-mode mixing is more complex than two-mode mixing, and the
general solution for a three mode eigensystem similar to equation
\ref{mix1} is sufficiently complicated that we do not attempt to
analyze it here. Instead, we wish to understand the effect of
rotational three-mode mixing between two core modes and an f-mode. For
instance, there are many modes that do not couple directly to f-modes,
but are coupled through an intermediary mode. One example is an $l=5$,
$m=-3$ s-mode mixed with a $l=4$, $m=-3$ t-mode, which in turn is
mixed with the $l=3$, $m=-3$ f-mode. We detail a solution of such a
three-mode system in \ref{threemode}. The main result is that
the $l=5$ s-mode can strongly mix with the f-mode even though they are 
not coupled directly. Thus, we expect that modes nearly degenerate
with an f-mode will obtain substantial gravitational potential
perturbations, even if they are not directly coupled with the f-mode.

We obtain the exact numerical solution to such a three-mode system and
plot the results in Figure \ref{3mode}. As expected, both core modes
undergo avoided crossings with the f-mode in which they mix strongly
with it. We also note that the strong mode mixing between the t-mode
and s-mode causes their frequencies to diverge away from one another
as the spin frequency is increased.\footnote{In a real system, these
  modes will also mix strongly with other s-modes and t-modes, which
  will mitigate this frequency divergence. Hence, it is important to
  extend mode mixing calculations to large values of $l$ in order to
  capture the realistic mixing behavior.}. This mode \textquotedblleft
repulsion" affects the location of avoided crossings with the f-mode,
causing them to occur at a spin frequency different than one would
expect from a two mode analysis. 
Finally, only the mode whose frequency is nearly degenerate with that of the f-mode is affected by
the avoided crossing.  Therefore, modes that differ substantially in
frequency from an f-mode will not strongly mix with it.

\subsection{Multi-Mode Rotational Mixing}
\label{multimodemix}

\begin{figure}
\begin{center}
\includegraphics[scale=0.4]{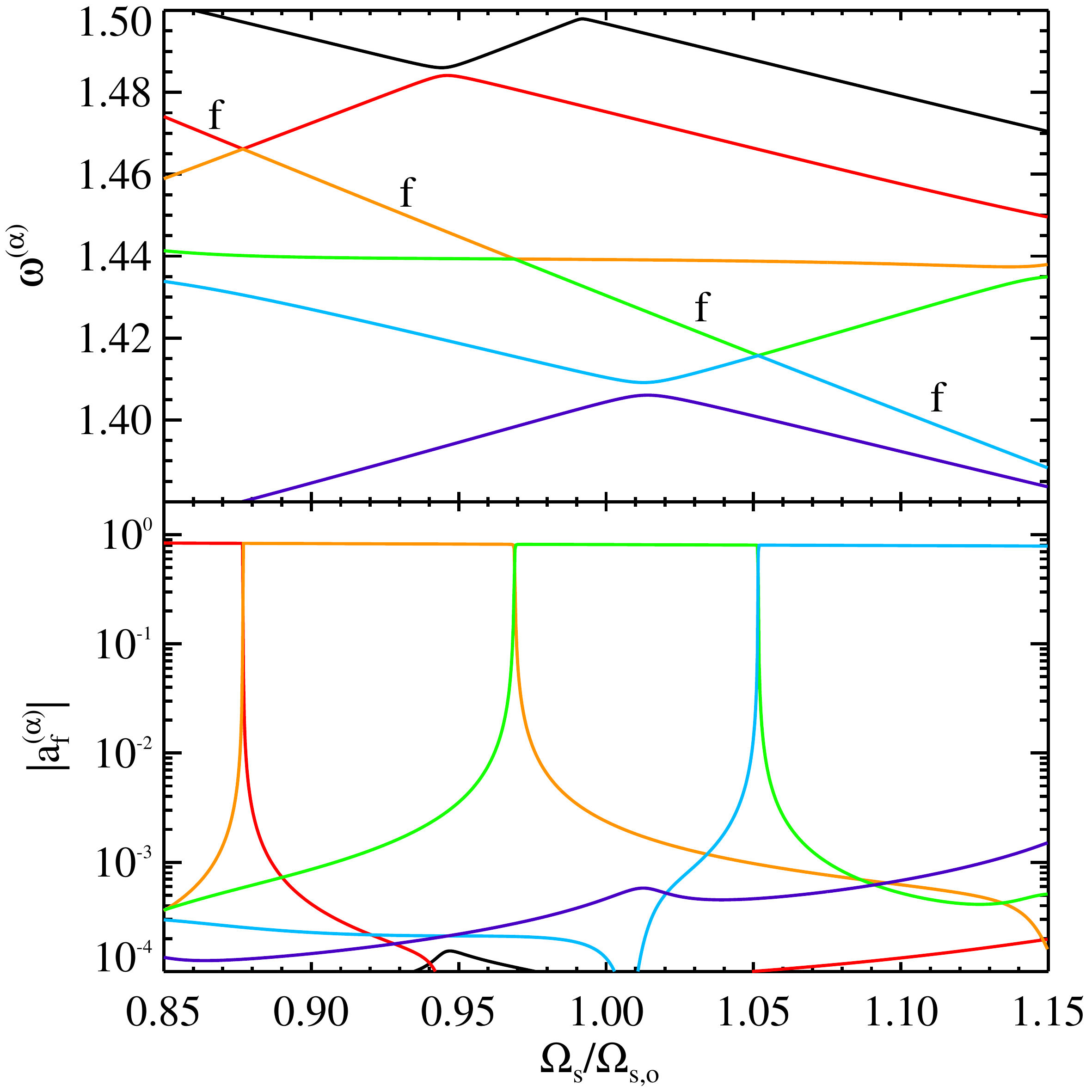}
\end{center} 
\caption{ \label{multimode} Same as Figure \ref{3mode}, but for a
  system containing 112 modes (only the six modes with frequency
  nearest the f-mode are shown). Here we plot the projection onto the
  f-mode, $a^{(\alpha)}_{\rm f} = b^{(\alpha)}_{\rm
    f}/\omega^{(\alpha)}$, with the modes normalized via equation
  (\ref{orth5}). The spin frequency $\Omega_s$ is plotted in units of
  Saturn's actual spin frequency. We have incorporated modes up to $l=18$, and have
  included all modes with frequency in the vicinity of the $l=3$
  f-mode. The \textquotedblleft f"-mode is labeled away from the avoided
  crossings.}
\end{figure}

We can now try to understand the effects of mode mixing in more realistic
systems, which may contain hundreds of coupled modes. To this end, we
solve equation (\ref{mat11}) for a system containing 112 modes for our
planetary model
with a fixed $\mu = 1.6 {\rm GPa}$ but with a range of spin frequencies.
We include all the modes with frequency comparable to the $l=3$
f-mode; this requires the inclusion of modes up to $l=18$. The results
are shown in Figure \ref{multimode}. The qualitative features of the
mode mixing are similar to the two and three-mode mixing shown in
Figures \ref{2mode} and \ref{3mode}. Although many modes are capable
of mixing with the f-mode, the small coupling coefficients $C_{{\rm f}\alpha}$
ensure that the width of each avoided crossing
is small compared to the spacing between modes. 
In other words, when many core modes are included, strong f-mode
mixing can occur at multiple values of the planetary spin. Since the
width of each strong mixing region is rather small, the chance of
finding a core mode strongly mixed with the f-mode is also very small.
A fine tuning in the spin frequency is required in order to realize
the observed fine ``splitting'' of f-modes in Saturn.

\section{Effect of Oscillation Modes on Saturn's Rings}
\label{gravpot}

Given a rotationally modified mode eigenfunction, we may calculate the
gravitational potential perturbation it produces exterior to the
planet. For a mode composed of a single spherical harmonic of angular
degree $l$ and azimuthal number $m$, the potential produced outside
the planet is
\beq
\delta \Phi_\alpha({\bf r},t) = A_\alpha \delta \Phi_\alpha(R) \bigg(\frac{R}{r}\bigg)^{l+1} Y_{lm}(\theta,\phi)e^{i\sigma_\alpha t},
\eeq
where $\delta \Phi_\alpha(R)$ is the potential perturbation at the
surface of the planet, calculated from our normalized mode
eigenfunction (equation \ref{norm}), $A_\alpha$ is the (unspecified)
amplitude of the mode, and $\sigma_\alpha$ is the mode frequency in
the inertial frame,
\beq
\sigma_\alpha = \omega_\alpha - m \Omega_s.
\eeq
This compact expression is modified in the presence of rotational
mixing because each mode acquires contributions from multiple angular
degrees $l$ (although $m$ remains a good ``quantum'' number). The
potential produced by a rotationally modified mode is then
\beq
\delta \Phi_\alpha({\bf r},t) = A_\alpha e^{i\sigma_\alpha t} \sum_{\beta} a^{(\alpha)}_{\beta} \delta \Phi_{\beta}(R) \bigg(\frac{R}{r}\bigg)^{l+1} Y_{l m}(\theta,\phi),
\eeq
where $a^{(\alpha)}_{\beta}$ is the projection of the rotationally
modified mode onto an original mode $\beta$ (determined via the method
in \ref{perturb}), and the sum is over all original modes
that contribute to the modified mode eigenfunction. For each term in
the sum, the value of $l$ corresponds to the mode $\beta$.

We are interested in the value of the potential $\delta \Phi_\alpha$
in the plane of the rings ($\theta=\pi/2$) at the location of the
outer Lindblad resonance. The resonance occurs where the forcing
frequency experienced by a particle is equal to the local epicyclic
frequency $\kappa$, i.e., where
\beq
\label{lindblad}
m(\Omega-\Omega_p) = \kappa,
\eeq
with $\Omega$ the local orbital frequency and $\Omega_p=-\sigma_\alpha/m$ 
the mode pattern frequency. 
The orbital and epicyclic frequencies are not exactly equal because of
the oblateness of Saturn, which causes the two frequencies to differ
by a factor $\sim 10^{-2}$ in the C-ring. Here we set them equal
because we are concerned primarily with the strength of the potential
(see Appenix \ref{rings} for a more accurate description). Then
the resonant location is
\beq
\label{r0}
r_L \simeq \bigg[\frac{(1-m)^2GM}{\sigma_\alpha^2}\bigg]^{1/3},
\eeq
and the gravitational potential at the resonant location in the ring is
\beq
\label{respot}
\delta \Phi_\alpha(r_L) \simeq A_\alpha e^{i\sigma_\alpha t + m \phi} \sum_{\beta} a^{(\alpha)}_{\beta}  Y_{lm}(\pi/2,0) \bigg[\frac{(1-m)^2}{\sigma_\alpha^2}\frac{GM}{R^3}\bigg]^{-(l+1)/3} \delta \Phi_{\beta}(R).
\eeq
The effective potential driving waves
at the Lindblad resonance is (Goldreich \& Tremaine 1979),
\begin{align}
\label{respot2}
\Psi_\alpha(r_L) &= \left[ \frac{d}{d \ln r} + \frac{2m\Omega}{\sigma_\alpha + m\Omega} \right] \delta \Phi_\alpha(r_L) \nonumber \\
& \simeq A_\alpha e^{i\sigma_\alpha t + m \phi} \sum_{\beta} a^{(\alpha)}_{\beta} W_{\beta} \delta \Phi_{\beta}(R)
\end{align}
where the dimensionless constant
\beq
W_\beta = (2m-l-1) Y_{lm}(\pi/2,0) \bigg[\frac{(1-m)^2}{\sigma_\alpha^2}\frac{GM}{R^3}\bigg]^{-(l+1)/3}.
\eeq
The surface density variation and the associated optical depth variation
produced near the Lindblad resonance are calculated in \ref{rings}.

The mode amplitudes $A_\alpha$ are unknown. To estimate them, we
assume that the most prominent $|m|=3$ wave measured by HN13 is produced by
an unmixed f-mode with frequency $\omega_3$, and we calculate the mode
amplitude $A_3$ required to generate the observed optical depth
variation in the ring (see \ref{rings}). We then assume the
modes follow energy equipartition, such that their amplitudes satisfy
$\omega_\alpha^2 |A_\alpha|^2 =
\omega_3^2|A_3|^2$.\footnote{Although energy equipartition may 
not apply to all modes, it is reasonable for our purpose
since the modes of interest have similar frequencies $\omega_\alpha$ and angular
  degrees $l$.} The typical mode amplitudes required to produce the
observed fluctuations are of order $|A| \sim 10^{-9}$, resulting in
radial surface displacements of order $\xi_r(R) \sim 30 {\rm cm}$ for
f-modes. These amplitudes are similar to those claimed to be observed
by Gaulme et al. (2011) in Jupiter and approximately what we might
expect if the modes are stochastically excited via convective
turbulence, analogous to the excitation of solar p-modes (see
discussion in Marley \& Porco 1993). However, the detailed dynamics of
mode excitation are beyond the scope of this paper.

\subsection{Comparison with Observations}

\begin{figure}
\begin{center}
\includegraphics[scale=0.45]{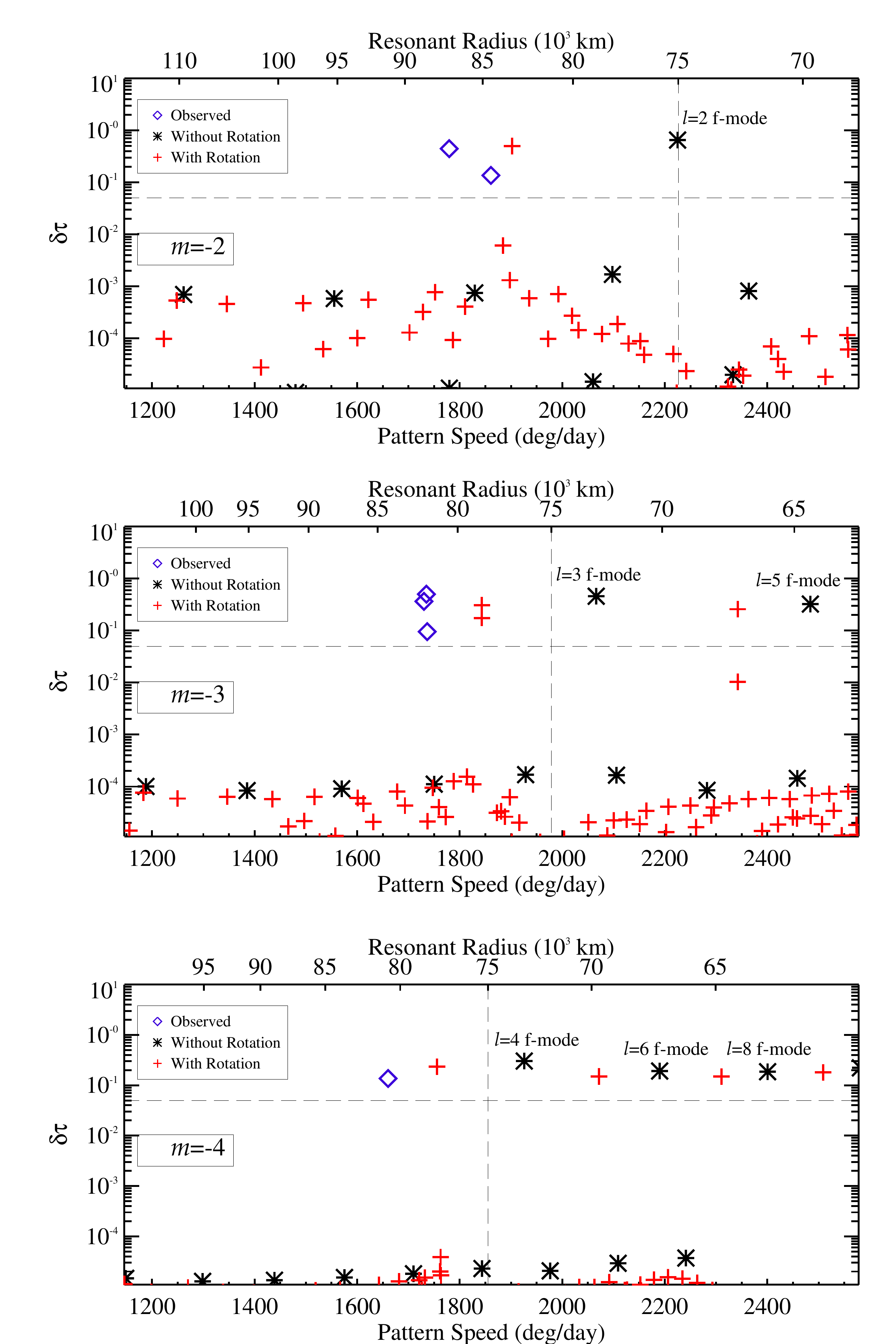}
\end{center} 
\caption{ \label{modesmu3} The predicted pattern frequencies
  $\sigma_\alpha/|m|$ and optical depth variations produced by waves
  at Lindblad resonances for the oscillation modes in our Saturn models,
  with $m=-2$ (top), $m=-3$ (middle), and $m=-4$ 
  (bottom). We have plotted modes of non-rotating models (black
  asterisks), models rotating uniformly at Saturn's observed rotation
  frequency (red pluses), and the observed waves (blue diamonds) from
  HN13. These plots are made for a Saturn model with $n=1$,
  $R_c=0.25$, $D=4$, and $\mu=1.6 {\rm GPa}$. The spin frequency has
  been slightly tuned ($\Omega_s/\Omega_{s,o} = 0.97$) to produce two
  strongly mixed modes near the f-mode in the middle panel. The
  vertical dashed line indicates the approximate inner edge of the
  C-ring, while the horizontal dashed line is the approximate minimum
  observable optical depth variation. Therefore, we should only expect
  to observe waves in the top-left corner of the figure. In the
  non-rotating models, the f-modes are the black asterisks in the top
  right, while the row of asterisks below them are the $l=|m|$
  s-modes.}
\end{figure}

\begin{figure}
\begin{center}
\includegraphics[scale=0.45]{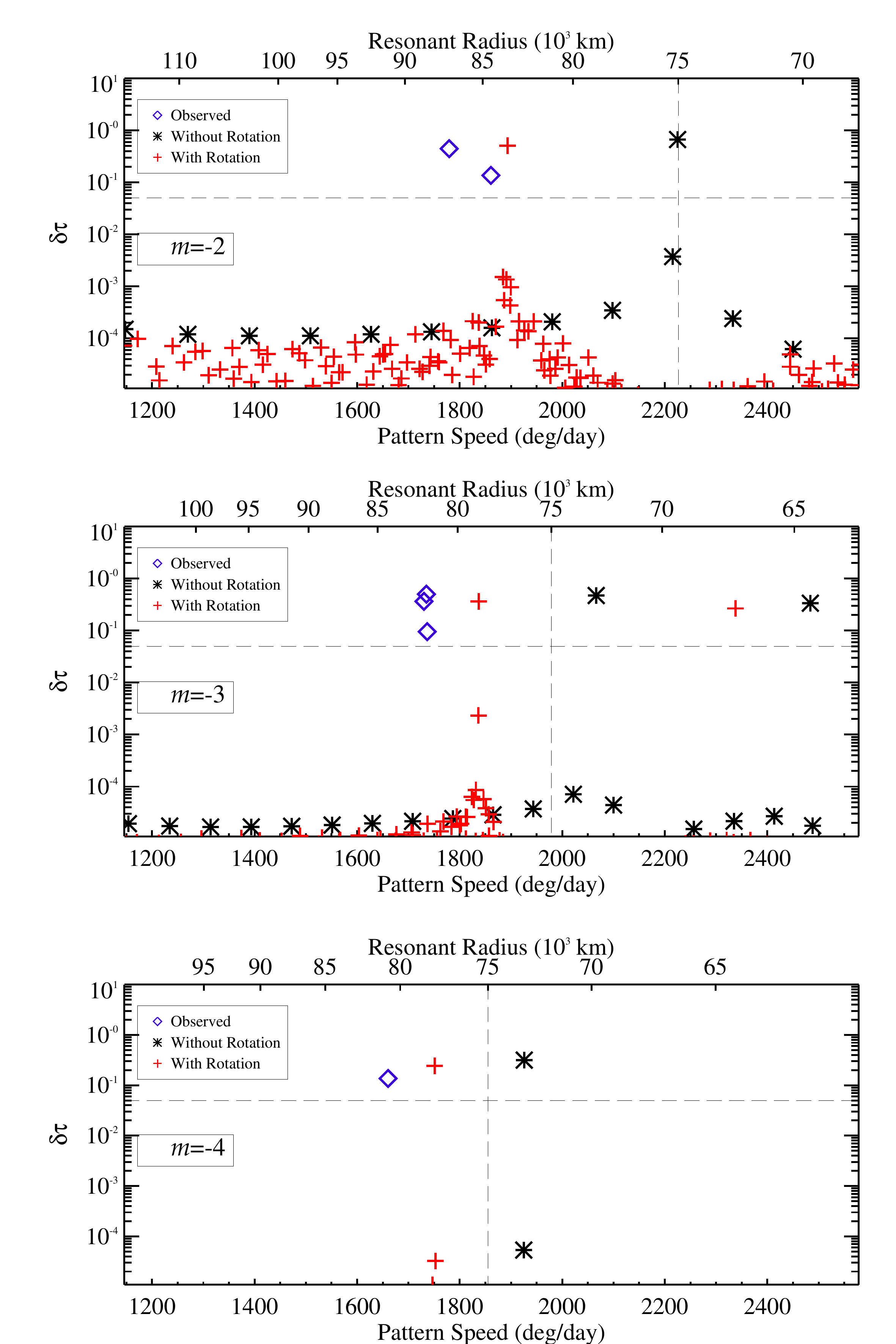}
\end{center} 
\caption{ \label{modesmu24} Same as Figure \ref{modesmu3} but for a
  Saturn model with $\mu=0.32 {\rm GPa}$. }
\end{figure}

\begin{figure}
\begin{center}
\includegraphics[scale=0.45]{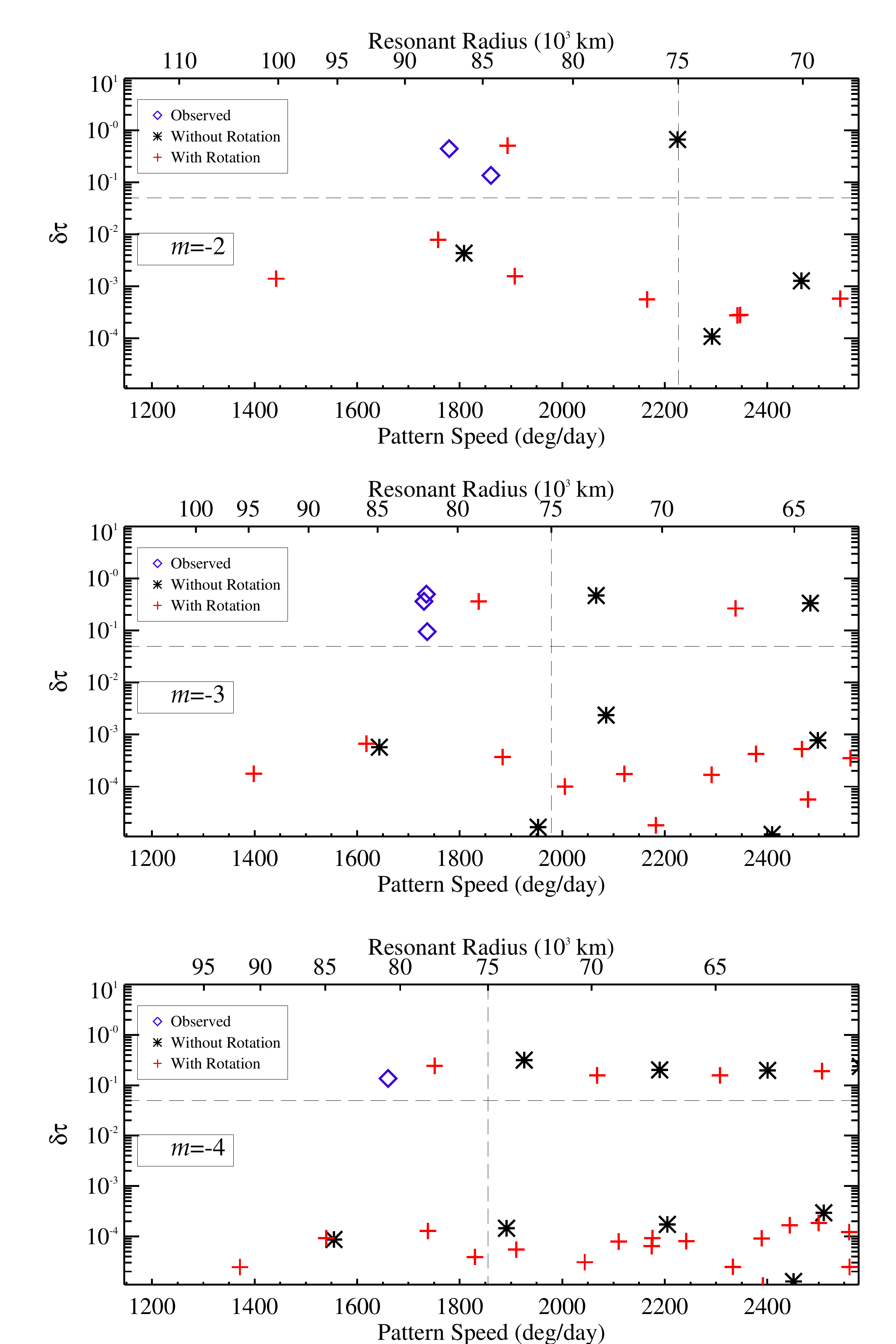}
\end{center} 
\caption{ \label{modesmu53} Same as Figure \ref{modesmu3} but for a
  Saturn model $\mu=8 {\rm GPa}$.}
\end{figure}

Figures \ref{modesmu3}-\ref{modesmu53} show the pattern frequencies
$\Omega_p$ (and the associated Lindblad radii $r_L$, calculated via
equation \ref{rL}) and the optical depth variations $\delta \tau$ (see
equation \ref{taupert}) produced by modes in our planetary models with
different values of the shear modulus. We have also plotted the
pattern frequencies and optical depth variations of the observed waves
in the C-ring as tabulated by HN13.

We begin by examining the mode pattern frequencies. In each model, the
pattern speed of the $l=-m$ f-mode in the non-rotating models is
consistently too large. Adding the Coriolis force lowers the pattern
frequency of the modified \textquotedblleft f"-mode (i.e., the lowest
frequency mode with a large value of $\delta \Phi_\alpha$), moving it
closer to the observed pattern frequencies, although the predicted
frequencies remain too high. Adding second-order rotational effects
is expected to lower the predicted frequencies further, making them more
consistent with observations. Obviously, with our simple planetary
models, we should not expect exact agreement. Nonetheless, the close
proximity of the predicted and observed pattern frequencies of the
\textquotedblleft f"-modes is strong evidence that the observed waves
are generated by Saturn's \textquotedblleft f"-modes. 

We can use Figures \ref{modesmu3}-\ref{modesmu53} to understand the
qualitative effects of rotationally-induced mode mixing. In the
absence of rotation, the spectrum of 
modes with significant gravitational perturbations is sparse, and is due almost
entirely to f-modes and low-order p-modes. However, as the rotation
rate increases, a core mode well-separated in frequency from the f-mode
obtains a potential perturbation
\beq
\label{dphi}
\delta \Phi_\alpha \approx \frac{|C_{{\rm f}\alpha}|}{\bar\omega_{\rm
    f}-\bar\omega_\alpha} \delta \Phi_{\rm f},
\eeq
where $\bar\omega_{\rm f}$ and $\bar\omega_\alpha$ are the rotationally
modified f-mode and core mode frequency, and $|C_{{\rm f}\alpha}|$ is
their coupling coefficient (equation \ref{Cdef}). Consequently, many
modes that had miniscule potential perturbations in the non-rotating
model (and thus fall below the plotted range of Figures
\ref{modesmu3}-\ref{modesmu53}) obtain larger potential perturbations
in the presence of rotation (although typically still orders of
magnitude smaller than the f-mode), causing the mode frequency
spectrum for rotating models to appear much denser in Figures
\ref{modesmu3}-\ref{modesmu53}.

A mode nearly degenerate with the f-mode will obtain a potential
perturbation $\delta \Phi_\alpha \simeq \delta \Phi_{\rm f}/\sqrt{2}$,
although in both modes are really hybrid f-modes. In this
case, the two modes are split in frequency by $\approx 2|C_{{\rm
    f}\alpha}|$. Our results suggest that such exact degeneracies are
rare in our planetary models, as described in Section \ref{multimodemix}. To
achieve significant mixing, the frequency separation between a core
mode and f-mode, $|\Delta_{\alpha \alpha'}|$, must be of order
$|C_{{\rm f}\alpha}|$. However, in our models, the typical frequency
separations $|\Delta_{\alpha\alpha'}|$ between core modes are much
larger than typical values of $|C_{{\rm f}\alpha}|$. 
Thus, the chances of a near degeneracy between a core mode and f-mode are small. 
Such degeneracies are possible (see e.g., the middle panel of Figure
\ref{modesmu3}), but usually require fine-tuning
of model parameters (such as the spin).
Since frequency splitting is observed by HN13 for both $m=-2$ and $m=-3$
modes, we conclude that it is unlikely to be produced solely by
Coriolis coupling between core modes and f-modes.

Finally, we examine the influence of the value of the shear modulus in
the core. If Saturn's solid core has a shear modulus $\mu \gtrsim 8
{\rm GPa}$ (Figure \ref{modesmu53}), then the core shear modes have
frequencies that are typically larger than the f-mode. Also, the core
mode spectrum is sparse, making it very unlikely for core modes to mix
strongly with the f-mode. Therefore, if $\mu \gtrsim 8 {\rm GPa}$, the
observed waves in Saturn's rings cannot be generated by the addition
of elastic core oscillation modes for reasonable Saturnian models.
On the other hand, if the shear modulus is $\mu \lesssim 1.6 {\rm GPa}$ (Figure
\ref{modesmu3}), there may exist many oscillation modes with
frequencies in the vicinity of the f-mode, allowing for the
possibility of strong mode mixing. However, a smaller shear
modulus also decreases the coupling coefficients, requiring a higher
degree of degeneracy (smaller $|\Delta_{\alpha \alpha'}|$) for
efficient mode mixing to occur.

\section{Discussion and Conclusions}
\label{discconc}

We have examined the influence of a solid core on the oscillation mode
spectrum of giant planets. In our planetary models, the rigidity of
the core has a rather small effect on the frequencies of f-modes or
low order p-modes, even for the largest core size ($R_c=0.25R$) and
the range of shear modulus $\mu$ considered in this paper: the mode
frequencies change change by less than $0.2$ percent.  However, the
addition of a solid core adds two new branches of oscillation modes:
the spheroidal and toroidal core modes, whose restoring force is the
elastic shear force of the solid material. The frequency spectrum of
these modes is dependent primarily on the core radius $R_c$ and shear
modulus $\mu$, with larger and less rigid (small $\mu$) cores
supporting dense spectra of low frequency shear modes. In the absence
of rotation, these modes are almost completely confined to the core,
and they produce negligible displacements or gravitational
perturbations at the surface of the planet.

We have also examined the influence of the Coriolis force on the mode
frequencies and eigenfunctions. In addition to changing the mode frequencies, 
the Coriolis force also induces mixing between oscillation modes of
equal azimuthal number $m$. Spheroidal modes of angular degree $l$ mix
with one another, as do toroidal modes. Furthermore, spheroidal modes
of angular degree $l$ mix with toroidal modes of angular degree
$l\pm1$, allowing for chains of mode mixing extending to arbitrary
values of $l$. Mode mixing is strongest when the rotationally modified
frequencies of two modes are nearly degenerate with one another. If
the value of the core shear modulus $\mu$ is small ($\mu \lesssim 3
{\rm GPa}$) but finite, the spectrum of core oscillation modes is
dense near the f-mode. Therefore, one or multiple core modes may be
nearly degenerate with the f-mode and mix strongly with it. These
strongly mixed modes would manifest themselves as hybrid
\textquotedblleft f"-modes, with nearly equal frequencies and similar
mode eigenfunctions.

One of the objectives of this investigation is to compare our planetary
oscillation calculations with the results of HN13, who have measured
the pattern numbers and frequencies of waves in Saturn's C ring that
appear to be excited by Saturn's oscillation modes. As speculated by
HN13 and originally proposed by Marley \& Porco (1993), we find that the
pattern frequencies and azimuthal numbers associated with individual
waves in Saturn's rings are consistent with being excited by Saturn's
prograde sectoral ($l=-m$) f-modes. It is possible that with more
realistic planetary models and including higher order spin affects,
the measured frequencies of these modes can be used to constrain
Saturn's interior structure.

HN13's observations of multiple wave trains, finely spaced in
frequency but with identical $m$, seem to indicate the existence of
multiple oscillation modes with frequencies near the frequencies of
Saturn's f-modes. We have shown that the rotation-induced mixing between shear modes in
Saturn's core and f-modes in Saturn's fluid envelope can in principle
lead to such finely spaced frequency splitting. Our theory has a number of
notable features when comparing with the observations:

(i) Our theory explains why the waves observed by HN13 all lie
close to the predicted locations of resonances with Saturn's f-modes.
We have shown that only core modes very near in frequency to the f-mode
will mix appreciably with it. Therefore, only these modes will obtain
large enough gravitational potential perturbations to produce
observable disturbances in Saturn's rings.

(ii) Our theory can in principle account for the relative magnitude of the observed frequency splittings (larger splitting for the $|m|=2$ modes, finer
splittings for the $|m|=3$ modes, and no observable splitting for the $|m|=4$ mode). 
Compared to higher degree modes, the $l=2$, $|m|=2$ f-mode penetrates
deeper into the planet. It has more inertia in the core, and thus
mixes more strongly with core modes, as evidenced by its larger
rotational coupling coefficients $|C_{{\rm f}\alpha}|$ with core
modes. Therefore we should expect that modes that mix strongly with the
$l=2$ f-mode to have larger frequency separations from it than the
modes that mix with the $l=3$, $|m|=3$ f-mode or the $l=4$, $|m|=4$ f-mode.

(iii) HN13's observations show that the finely spaced waves in Saturn's ring have  
different optical depth variations. 
In our theory, modes nearly degenerate with the f-mode may obtain
potential perturbations comparable to (but less than) that of the
unperturbed f-mode, with the precise values determined by the
rotational coupling coefficient and degree of frequency detuning.


Despite these positive features, our theory of rotation-induced mode
mixing suffers a serious fine-tuning problem when explaining the HN13
observations.  Our calculations indicate that f-modes mix with core
shear modes efficiently only when their frequency differences are very
small. Such small frequency detunings are unlikely to occur unless
the planetary model (including the rotation rate) is fine-tuned to
produce a core mode nearly degenerate with the f-mode.  Even with a
finely tuned model, it is difficult to simultaneously achieve
the observed $|m|=2$ and $|m|=3$ frequency splittings.  Another
concern is that a very small value of the shear modulus of the solid
core ($\mu \lesssim 3 \ {\rm GPa}$) is required for our mechanism 
(in order to produce a dense spectrum of core modes near the f-mode
frequency).
In comparison, the shear modulus of Earth's core is $\mu \approx 100
\ {\rm GPa}$. (Laio et al. 2000), and the pressure in Saturn's core is
of order $P \sim 2 \ {\rm TPa}$. Although such a small value of the
shear modulus may seem unlikely, it may be produced by the high
pressure superionic phase of ice (see references in Section
\ref{shearmod}) likely to exist in Saturn's core.

There are a number of effects or perturbations that we have neglected
in our paper and are worthy of further study.  In general, for any
perturbation ${\cal P}$ to induce strong mixing between an f-mode and
another mode $\alpha$, the mixing matrix element $\calP_{{\rm f}\alpha}$ 
must satisfy (see Section \ref{rotationcoup})
\beq 
|\calP_{{\rm f}\alpha}|\gtrsim |\bar\omega_f-\bar\omega_\alpha|,
\eeq
where $\bar\omega_f$ and $\bar\omega_\alpha$ are the mode frequencies
(including the ``self''-correction due the perturbation).

(i) We have already noted (see Section \ref{intro}) that higher-order rotational effects 
(including the rotational distortion of the planet) can significantly change the
f-mode frequency. This self-correction is of order 
$\sim \Omega_s^2/\omega_{\rm f}$, and should be included when making detailed
predictions of f-mode frequencies in different planetary models. 
The related mixing element $|\calP_{{\rm f}\alpha}|$
is of the same order or 
less than $\Omega_s^2/\omega_f$, but nevertheless can be much larger compared
to $|C_{{\rm f}\alpha}|$ associated with the Coriolis term. This may 
help alleviate the fine-tuning problem.

(ii) Differential rotation and other internal flows (such as that associated with
slow convective motion) can also change mode frequencies, by the amount
$\sim \Delta\Omega_d$ (differential rotation) or $\sim V_{\rm bulk}/R$ (for 
bulk flow speed of order $V_{\rm bulk}$). The related mixing element is
of the same order or less. Since the observed fine splittings in HN13 
are a few percent or less, these effects cannot be neglected.

(iii) Internal planetary magnetic fields can change the f-mode frequency by the amount
$\sim (v_A/R)^2/\omega_{\rm f}$, where $v_A=\sqrt{B^2/4\pi\rho}$
is the charateristic Alfven speed. The fractional correction,
$\sim (v_A/R\omega_{\rm f})^2$ is less than $10^{-10}$ even for an extremely strong
($B\sim 1$~kG) internal field. This effect may be neglected.

(iv) In this paper we have focused on f-mode mixing with shear modes
in the core.  But there may be other types of modes that can
participate in the mixing.  We have noted before (Section 1) that
inertial modes or Rossby modes cannot directly mix with the f-mode
because of their frequency mismatch. But they may be involved in
secondary mixing in a multiple-mode mixing system (see Section \ref{multimodemix}).
Also, in the presence of stratification of heavy elements (see Leconte
\& Chabrier 2012), the fluid envelope may support gravity modes. These
new modes can increase the ``phase space'' of mixing with f-modes,
allowing them to be observed indirectly.

Overall, it remains unclear at present whether the finely spaced
density waves observed in Saturn's C-ring can be naturally produced by
the simultaneous presence of finely split oscillation modes inside
Saturn. In this regard, we should keep in mind the possibility that
the observed waves are produced at different times, or the different
wave frequencies reflect secular changes in the mode properties (such as amplitude and frequency) or internal properties (such as
differential rotation) of the planet.

\section*{Acknowledgments} 

We thank Matt Hedman and Phil Nicholson for extensive discussions
regarding the Cassini data used to measure the properties of the waves
in Saturn's rings. This work has been supported in part by NSF grants
AST-1008245, 1211061, NASA grants NNX12AF85G and NNX10AP19G.

\appendix

\section{Elastic Oscillation Equations}
\label{app1}

In this work we consider purely adiabatic and elastic oscillations,
i.e., we neglect damping produced by non-adiabaticity and
anelasticity. Under these approximations, the elastic forces act like
a spring, which is characterized by the stress tensor
\beq
\boldsymbol{\sigma}= K (\bnab \cdot \bxi) {\bf I} + 2 \mu {\bf s}.
\eeq
Here, $K$ is the adiabatic bulk modulus, related to the pressure $P$
via $K=\Gamma_1 P$, with $\Gamma_1$ the adiabatic index of the
material. Also, $\mu$ is the shear modulus, $\bxi$ is the vector
displacement, ${\bf I}$ is the identity matrix, and the deviatoric
strain tensor ${\bf s}$ is defined as
\beq
{\bf s} \equiv \frac{1}{2} \Big[\bnab \bxi + (\bnab \bxi)^T\Big] - \frac{1}{3}\big(\bnab \cdot \bxi \big) {\bf I} 
\eeq

We assume that the unperturbed state of the planet has zero elastic stress. The perturbed hydroelastic oscillation equation reads
\beq
\label{osceq}
\frac{\partial^2\boldsymbol{\xi}}{\partial t^2} = \frac{\delta\rho}{\rho^2}\bnab P - \frac{1}{\rho}\bnab\delta P -\bnab\delta\Phi + \delta\textbf{f}_e,
\eeq
where the perturbed elastic force $\delta\textbf{f}_e$ is
\begin{align}
\delta\textbf{f}_e &= \frac{2}{\rho}\bnab\cdot\left(\mu{\bf s}\right) \\ \nonumber
&= {2\over \rho}\bnab\mu\cdot{\bf s} + \frac{\mu}{\rho}\left(\nabla^2\bxi+\frac{1}{3}\bnab(\bnab\cdot\bxi)\right)
\label{eq2}
\end{align}

We decompose all variables into spherical harmonics, or spherical vector harmonics where appropriate:
\begin{equation}
\begin{split}
\boldsymbol{\xi}(\textbf{r},t) &= \sum_{l,m} \left[U_l(r) Y_{lm} {\bf \hat{r}} + V_l(r) r\bnab Y_{lm} + W_l(r)\bnab\times(\textbf{r} Y_{lm})\right] e^{i\omega t},\\
\delta\Phi (\textbf{r},t) &= \sum_{l,m} \left[\delta\Phi_{l}(r) Y_{lm}\right]e^{i\omega t},
\end{split}
\end{equation}
and similarly for $\delta\rho$ and $\delta P$.
Then the equations become separated such that only r-dependence is
left, and the equations for different harmonics are
self-contained. From now on we will drop the subscript $l$ and 
consider the equations for a single harmonic. Explicitly, the
displacement vector becomes:
\begin{equation}
\begin{split}
\xi_r &= U(r) Y_{lm}\\
\xi_\theta &= V(r) \frac{\partial Y_{lm}}{\partial\theta} + \frac{W(r)}{\sin\theta}\frac{\partial Y_{lm}}{\partial\phi}\\
\xi_\phi &= \frac{V(r)}{\sin\theta}\frac{\partial Y_{lm}}{\partial\phi}-W(r) \frac{\partial Y_{lm}}{\partial\theta}.
\end{split}
\label{xi}
\end{equation}
In addition, there is the Poisson equation for $\delta\Phi$:
\begin{equation}
\nabla^2 \delta\Phi = 4\pi G\delta\rho.
\label{poisson}
\end{equation}
We also define the function $\alpha(r)$ via
\beq
\alpha Y_{lm} = \bnab \cdot \bxi = \left[ \dot{U} + \frac{2}{r} U - \frac{l(l+1)}{r} V \right] Y_{lm},
\eeq
where the dot denotes radial derivative. The continuity equation gives
\begin{align}
\delta \rho &= - \bxi \cdot \bnab \rho - \rho \bnab \cdot \bxi \nonumber \\
 & = - \left( U \dot{\rho} + \rho \alpha \right) Y_{lm}.
\end{align}

Equation (\ref{osceq}) represents three 2nd-order ordinary differential
equations for $U$, $V$, $W$, while equation (\ref{poisson}) is a 2nd-order 
ODE for $\delta \Phi$. We wish to transform these four
second-order ODEs into eight first-order ODEs. First we define a new
quantity, the Lagrangian traction $\textbf{T}\equiv{\bf
  \hat{r}}\cdot\Delta\boldsymbol{\sigma}$, where $\Delta\sigma_{ij} =
K\alpha\delta_{ij}+2\mu s_{ij}$. Then we have
\begin{align}
\textbf{T} &= {\bf \hat{r}}\cdot\Delta\boldsymbol{\sigma}\nonumber\\
&= T_r(r) Y_{lm} {\bf \hat{r}}+ T_{\perp}(r)r\bnab Y_{lm}+T_{t}(r)\bnab\times(\textbf{r}Y_{lm})\nonumber\\
&= \left( \left[K-\frac{2}{3}\mu \right] \alpha+2\mu \dot{U} \right)Y_{lm}{\bf \hat{r}} \nonumber\\
&+\mu\left\{\left(\frac{U}{r}+\dot{V}-{V\over r}\right)\frac{\partial Y_{lm}}{\partial\theta}+\left(\dot{W}-{W\over r}\right)\frac{1}{\sin\theta}\frac{\partial Y_{lm}}{\partial\phi}\right\}\hat{\theta}\nonumber\\
&+\mu\left\{\left(\frac{U}{r}+\dot{V}-\frac{V}{r}\right)\frac{1}{\sin\theta}\frac{\partial Y_{lm}}{\partial\phi}-\left(\dot{W}-{W\over r}\right)\frac{\partial Y_{lm}}{\partial\theta}\right\}\hat{\phi}.
\label{traction}
\end{align}
We now define 8 new variables as follows:
\begin{align}
y_1 &= U,\\
y_2 &\equiv T_r = \bigg(K - \frac{2}{3}\mu\bigg)\alpha + 2\mu \dot{U},\\
y_3 &= V,\\
y_4 &\equiv T_\perp = \mu\bigg(\frac{U}{r} + \dot{V} - \frac{V}{r}\bigg),\\
y_5 &= \delta\Phi,\\
y_6 &= \dot{y}_5 + 4\pi G\rho y_1,\\
y_7 &= W,\\
y_8 &\equiv T_t = \mu\bigg(\dot{W} - \frac{W}{r}\bigg).
\end{align}
\label{vardefs}
Note that $y_1,y_3,y_7$ are the displacement components, while
$y_2,y_4,y_8$ are the Lagrangian tractions. $y_6$ is the Lagrangian
gravitational attraction, chosen as a variable for convenience when
applying boundary conditions (see section \ref{bc}). The final
oscillation equations are:
\begin{align}
\dot{y}_1 &= -{2(K-{2\over 3}\mu)\over(K+{4\over 3}\mu)}{y_1\over r} + {1\over(K+{4\over 3}\mu)}y_2 + {l(l+1)(K-{2\over 3}\mu)\over(K+{4\over 3}\mu)}{y_3\over r},\label{eq:eqy1}\\
\dot{y}_2 &= \left[-4\rho g -\omega^2\rho r +{12\mu K\over(K+{4\over 3}\mu)r}\right]{y_1\over r} - {4\mu\over(K+{4\over 3}\mu)}{y_2\over r}\nonumber \\
&+l(l+1)\left[\rho g-{6\mu K\over (K+{4\over 3}\mu)r}\right]{y_3\over r} + l(l+1){y_4\over r} + \rho y_6,\\
\dot{y}_3 &= -{y_1\over r} + {y_3\over r} + {y_4\over\mu},\\
\dot{y}_4 &= \left[\rho g - {6K\mu\over(K+{4\over 3}\mu)r}\right]{y_1\over r}-{(K-{2\over 3}\mu)\over(K+{4\over 3}\mu)}{y_2\over r}\nonumber \\
&+\left\{-\omega^2\rho r + {2\mu[(K-{2\over 3}\mu)(2l^2+2l-1)+2\mu(l^2+l-1)]\over(K+{4\over 3}\mu)r}\right\}{y_3\over r}\nonumber \\
 &- 3{y_4\over r} + \rho{y_5\over r},\\
\dot{y}_5 &= -4\pi G\rho y_1 + y_6,\\
\dot{y}_6 &= 4\pi G\rho l(l+1){y_3\over r} + {l(l+1)\over r}{y_5\over r} - {2\over r} y_6,\\
\dot{y}_7 &= {y_7\over r} + {y_8\over\mu},\\
\dot{y}_8 &= \left[-\omega^2\rho r + {\mu(l^2+l-2)\over r}\right]{y_7\over r} - {3\over r}y_8.
\label{vareqs}
\end{align}
Note that $y_7$ and $y_8$ completely decouple from the other six
equations. The first six equations are integrated to obtain the
spheroidal modes, while the last two are integrated to obtain the
toroidal modes.

In fluid regions where $\mu \rightarrow 0$, $y_4 \rightarrow 0$ and
$y_3$ is found from the algebraic relation
\beq
y_3 = \frac{1}{r \omega^2} \bigg( g y_1 - \frac{1}{\rho} y_2 +  y_5 \bigg).
\eeq
Thus only four ODEs are needed for the fluid region.

\subsection{Boundary Conditions}
\label{bc}

The boundary conditions (BCs) for elastic oscillations can be found in
Dahlen and Tromp (1998) and in Crossley (1975). For the spheroidal
oscillations, the BCs at $r=0$ are, to lowest order in $r$,
\begin{align}
y_1 &= A r^{l-1},\label{eq:bcin1} \\
y_2 &= 2(l-1)\mu A r^{l-2}, \\
y_3 &= \frac{A}{l} r^{l-1}, \\
y_4 &= \frac{2(l-1)\mu A}{l} r^{l-2}, \\
y_5 &= \bigg[-\frac{4\pi G \rho A}{l} + \frac{B}{l}\bigg] r^l. \\
y_6 &= B r^{l-1}. \label{eq:bcin6}
\end{align}
For $l=0$, $A=0$ and higher order terms in $r$ are needed to establish the BCs (see Crossley 1975). When $l=1$, the displacements are finite at the center of the planet. Note that for $l=2$, the values of the tractions $T_r$
and $T_\perp$ are finite, i.e., the core is undisplaced but is under
stress. For $l>2$, the values of all perturbation variables are zero
at the center of the planet. 
Equations (\ref{eq:bcin1})-(\ref{eq:bcin6})
comprise only two independent BCs (corresponding to the undetermined constants $A$ and
$B$), and the additional inner BCs are trivially satisfied due to the form
of the oscillation equations. In practice, for $l \geq 2$, we use the
four inner BCs
\begin{align}
\label{BCin2}
y_1 &= 0,\\
y_2 &= l y_4,\\
y_4 &= 0,\\
y_6 &= 0,
\end{align}
although other choices are possible.

The three independent boundary conditions at the surface of the planet are 
\begin{align}
y_2 &= 0,\\
y_4 &= 0,\\
y_6 &= -\frac{l+1}{R} y_5.
\end{align}
Note that in the fluid envelope, the first of these conditions is
equivalent to the boundary condition $\Delta P=0$ commonly used in
asteroseismology. The condition on $y_4$ is trivially satisfied in the
planetary model with a fluid envelope, since $y_4=0$ in fluid
regions. In our calculations, we instead use the outer BCs involving
$y_2$ and $y_6$, the four inner BCs listed above, and the jump
conditions at core-envelope interface described below.

At the solid core-fluid envelope interface, the values of $y_1$, $y_2$,
$y_4$, $y_5$, and $y_6$ must be continuous. There is no relation to
determine the change in $y_3$ across 
the interface, and in general $y_3$ is discontinuous.
Note that the continuity of $y_6$ across an interface implies a discontinuity in the
gravitational perturbation, $\delta d\Phi/dr$. Also note that the
continuity of $y_4$ implies that it is zero at a fluid-solid
interface. For torsional oscillations of $l\geq2$, the BCs are $y_7=0$ at $r=0$
and $y_8=0$ at $r=R$. At a fluid-solid interface, $y_8$ is continuous
but $y_7$ is generally discontinuous. 

Solving the eigensystem (\ref{eq:eqy1})-(\ref{vareqs}) for the eigenvalue $\omega$ also
requires a normalization boundary condition. Typically, the normalization
$y_1(R)=1$ is used, although the choice of
normalization is entirely arbitrary. 
For oscillation modes in planets
with a solid core and a fluid envelope, 
this condition is poorly suited for numerical computation. The reason is
that p-modes confined to the fluid envelope have very large relative
surface displacements, while shear modes confined to the solid core
have very small relative surface displacements. 
For our models, we find the normalization condition $y_3=1$
just below the core-envelope boundary allows our code to quickly
converge for both p-modes and shear modes. The eigenmodes are then
renormalized via equation (\ref{norm}).

\section{Solving for Rotationally Mixed Modes}
\label{perturb}

In the rotating frame of the planet, the perturbed momentum equation takes the form
\beq
\label{momeq}
-\omega^2 \bxi + 2 i \omega \boldsymbol{\Omega}_s \times \bxi + \mathcal{H} \bxi = 0,
\eeq
where $\mathcal{H} \bxi$ is given by the negative of the right hand
side of equation (\ref{osceq}), and we have used $\bxi \propto e^{i\omega t}$. Let 
\beq
\label{xidecomp}
\bxi({\bf r}) = \sum_\alpha a_\alpha \bxi_\alpha ({\bf r}),
\eeq
where $\bxi_\alpha({\bf r})$ satisfies
\beq
\label{eigen}
-\omega_\alpha^2 \bxi_\alpha + \mathcal{H} \bxi_\alpha = 0.
\eeq
Then we obtain
\beq
\label{eigen2}
\left( \omega_\alpha^2-\omega^2 \right) a_\alpha + 2 \omega \sum_\beta C_{\alpha \beta} a_\beta = 0,
\eeq
where 
\begin{align}
\label{Cab}
C_{\alpha \beta} &= \langle \bxi_\alpha \vert i \boldsymbol{\Omega}_s \times \bxi_\beta \rangle \nonumber \\
&= \int dV \rho \bxi_\alpha^* \cdot \left(i \boldsymbol{\Omega}_s \times \bxi_\beta \right).
\end{align}
Explicit expressions for the mixing coefficient $C_{\alpha\beta}$ are given in  \ref{rotcoup}. 

Equation (\ref{eigen2}) represents an eigensystem for the modified
eigenvalues $\omega$, with each row in the matrix equation indexed by
$\alpha$ and each column indexed by $\beta$. For each eigenvalue
$\omega^2$, the complex components of the eigenvector ${\bf a} =
\{a_{1},...a_N\}$ represent the projection of the new eigenmode onto
each original mode $\beta$. In matrix form, equation (\ref{eigen2}) is a
quadratic eigenvalue problem:
\beq
\label{mat5}
\big( \boldsymbol{\Omega} + 2 \omega \bf{C}\big) {\bf a} = \omega^2 \bf{I} {\bf a},
\eeq
where $\boldsymbol{\Omega}={\rm diag}\{\omega_1^2,...,\omega_N^2\}$,
${\bf I}$ is the identity matrix, and
\beq
{\bf C} = \begin{bmatrix} C_{11} & \cdots & C_{1N} \\ 
\vdots & \ddots & \vdots \\
 C_{N1} & \cdots & C_{NN} 
\end{bmatrix}
\eeq
The $N\times N$ eigenvalue problem of equation (\ref{mat5}) is
equivalent to the $2N\times 2N$ eigenvalue problem
\beq
\label{mat6}
\begin{bmatrix} 0 & {\bf I} \\ \boldsymbol{\Omega} & 2{\bf C} \end{bmatrix} \left[ \begin{array}{c} {\bf a} \\ {\bf b} \end{array} \right] = \omega \left[ \begin{array}{c} {\bf a}  \\ {\bf b} \end{array} \right],
\eeq
where ${\bf b} = \omega {\bf a}$. Equation (\ref{mat6}) is equivalent to
the phase space mode expansion, where a mode ${\bf Z}$ of eigenvalue
$\omega$ is defined by both its displacement and velocity vectors such
that ${\bf Z}=[{\bf a},{\bf b}]=[{\bf a},{\bf \omega a}]$. Solving the
eigensystem of equation (\ref{mat6}) yields $2N$ eigenvalues
$\omega$. In the limit $\Omega_s \rightarrow 0$, the solutions come in pairs,
 $\omega_\alpha$ and $-\omega_\alpha$, 
representing the prograde and retrograde modes. Note that in this
form, the matrix on the left hand side of equation (\ref{mat6}) is not
explicitly Hermitian, however, we show below that the system can be
written in a Hermitian form.

We solve the eigensystem \ref{mat6} for the eigenvalues $\omega$ and
the associated eigenfunctions ${\bf a}$. Because only modes of the
same $m$ are coupled (see Section \ref{rotcoup}), we may solve
equation \ref{mat6} separately for each value of $m$ we wish to
consider. For negative $m$, the positive values of $\omega$ represent
prograde modes, and the negative values of $\omega$ represent
retrograde modes. In principle, we must include every oscillation mode
(i.e., for a given $m$ we must include modes of all $l$ and $n$) in
the eigensystem. In practice, we truncate the eigensystem at
finite values of $l$ and $n$. To determine which modes to include, we
can solve equation (\ref{mat6}) for a limited set of modes near the
$l=|m|$
f-modes, and then extend our calculations to larger values of
$l$ and larger frequency ranges to see if the results change. For the
planetary models, spin frequencies, and mode frequencies considered in
this paper, we find that including only modes with $l \lesssim 14$ and
$\omega_{l,n}/3 \lesssim \omega_f \lesssim 3 \omega_{l,n}$ yields a
good approximation.

\subsection{Alternative Formalism}

The eigensystem equation (\ref{mat6}) can be solved more elegantly, as
shown in Dyson \& Schutz (1979) and DT98. We seek to solve the
eigenvalue problem
\beq
\label{sys}
\Big[ \mathcal{H} + \omega \bar{\mathcal{C}} \Big] \bxi = \omega^2 \bxi,
\eeq
where $\mathcal{H}$ has the same definition as above, and
$\bar{\mathcal{C}}=2i{\bf \Omega}_s \times$.  Defining each
eigenvector via its six-dimensional eigenfunction
\beq
\label{mat7}
{\bf Z} = \left[ \begin{array}{c} {\bxi} \\ \omega {\bxi} \end{array} \right].
\eeq
We rewrite equation (\ref{sys}) as 
\beq
\label{mat8}
\Bigg( \begin{bmatrix} 0 & \mathcal{I} \\ \mathcal{H} & 0 \end{bmatrix} + \begin{bmatrix} 0 & 0 \\ 0 & \bar{\mathcal{C}} \end{bmatrix} \Bigg) {\bf Z} = \omega {\bf Z},
\eeq
where $\mathcal{I}$ is the identity operator. We decompose ${\bf Z}$ in
terms of original eigenvectors,
\beq
\label{xidecomp}
{\bf Z} = \sum_\beta a_\beta {\bf Z}_\beta,
\eeq
where here the sum runs over the $2N$ components $\beta$ (accounting
for both negative and positive eigenfrequencies) because we have
employed the phase space mode expansion, in contrast to the
configuration space expansion used in equation \ref{xidecomp}. Recalling
that $\mathcal{H} \bxi_\beta = \omega_\beta^2 \bxi_\beta$, equation
(\ref{mat8}) becomes
\beq
\label{mat9}
\sum_\beta \omega_\beta a_\beta {\bf Z}_\beta + \sum_\beta \begin{bmatrix} 0 & 0 \\ 0 & \bar{\mathcal{C}} \end{bmatrix} a_\beta {\bf Z}_\beta = \sum_\beta \omega a_\beta {\bf Z}_\beta.
\eeq
We now multiply by the auxiliary operator 
\beq
\label{P}
\mathcal{P} =  \begin{bmatrix} \mathcal{H} & 0 \\ 0 & \mathcal{I} \end{bmatrix},
\eeq
and take the inner product with ${\bf Z}_\alpha$ to obtain
\beq
2 \omega_\alpha^3 a_\alpha + 2\omega_\alpha \sum_\beta \omega_\beta a_\beta C_{\alpha\beta} = 2 \omega_\alpha^2 \omega a_\alpha,
\eeq
where $C_{\alpha\beta}$ has the same definition as above. Defining $b_\alpha = \omega_\alpha a_\alpha$, we have
\beq
\label{mat11}
\omega_\alpha b_\alpha + \sum_\beta C_{\alpha\beta} b_\beta =  \omega b_\beta.
\eeq
Equation (\ref{mat11}) represents a Hermitian eigensystem because
$C_{\alpha\beta} = C_{\beta\alpha}^*$, therefore it is amenable to
numerical matrix solving techniques. Once the eigenvectors ${\bf b}$
are determined, components of the desired eigenvector ${\bf a}$ are
obtained by $a_\beta = b_\beta /\omega_\beta$.

\subsection{Rotational Mixing Coefficients}
\label{rotcoup}

The element $C_{\alpha\alpha'}$ of the rotational coupling matrix is defined as
\beq
\label{Cdef}
C_{\alpha\alpha'} = i \Omega_s \int dV \rho \bxi_{\alpha}^* \cdot \big({\bf \hat{z}} \times \bxi_{\alpha'}\big),
\eeq
where ${\bf \hat{z}}$ is the unit vector in the $z$ direction, and the
integral is over the volume of the planet. Explicitly, the value of
$C_{\alpha\alpha'}$ is (see also Dahlen \& Tromp 1998)
\begin{align}
\label{C2}
C_{\alpha\alpha'} &=  m \Omega_s \delta_{ll'} \delta_{mm'} \int^R_0 dr \rho r^2 \Big( UV'+VU'+VV'+WW' \Big) \nonumber \\ 
&- \frac{i\Omega_s}{2} \big( S_{lm}\delta_{ll'+1} + S_{l'm}\delta_{ll'-1}\big)\delta_{mm'} \nonumber \\
&\times \int^R_0 dr \rho r^2 \bigg[ \Big(k_l^2-k_{l'}^2-2\Big) UW' + \Big(k_l^2-k_{l'}^2+2\Big) U'W  + \Big(k_l^2+k_{l'}^2-2\Big) \Big(VW'-V'W\Big)\bigg],
\end{align}
with $k_l^2 = l(l+1)$ and 
\beq
S_{lm} = \bigg[\frac{(l+m)(l-m)}{(2l+1)(2l-1)}\bigg]^{1/2}.
\eeq
For compactness, we have dropped the $\alpha$ subscript on the right
hand side of equation (\ref{C2}). The displacement functions $U$, $V$,
and $W$ are defined in equation (\ref{xi}). The first integral in
equation (\ref{C2}) accounts for spheroidal-spheroidal coupling and
toroidal-toroidal mode coupling. For $\alpha=\alpha'$, it reduces to
the conventional rotational splitting parameter. The second integral
in equation (\ref{C2}) accounts for spheroidal-toroidal mode
coupling. Note the Hermitian nature of $C_{\alpha\alpha'}$, i.e.,
$C_{\alpha\alpha'} = C^*_{\alpha'\alpha}$.

The value of $C_{\alpha\alpha'}$ is zero unless the modes satisfy
certain angular selection rules. In particular, only modes of equal
azimuthal number $m=m'$ can couple since the introduction of rotation
does not break the axial symmetry of the problem. Similarly,
spheroidal modes couple only to other spheroidal modes of equal
angular degree such that $l=l'$. The same is true for toroidal
modes. However, spheroidal modes of degree $l$ may couple to toroidal
modes of degree $l'=l\pm1$, and vice versa. Therefore, spheroidal
modes of $Y_{l,m}$ are coupled to toroidal modes of $Y_{l+1,m}$, which
in turn are coupled to spheroidal modes of $Y_{l+2,m}$, etc. Thus, the
inclusion of the Coriolis forces introduces an infinite chain of
coupling with modes of higher angular degree $l$.

\subsection{Mode Normalization}

The modes obtained by solving the eigensystem (\ref{mat6}) or
(\ref{mat11}) must be appropriately normalized, because they are no
longer orthonormal under equation (\ref{norm}). Instead, the modes
satisfy the modified orthonormality condition
\beq
\label{orth2}
\langle\langle {\bf Z}^{(\alpha)} \vert {\bf Z}^{(\alpha')} \rangle\rangle \equiv \langle {\bf Z}^{(\alpha)} \vert \mathcal{P} \vert {\bf Z}^{(\alpha')} \rangle = 2 \big(\omega^{(\alpha)} \big)^2 \delta_{\alpha,\alpha'},
\eeq
with $\mathcal{P}$ defined by equation (\ref{P}), and the $(\alpha)$ superscript is the index of the rotationally modified mode. Recall that the decomposition of equation (\ref{xidecomp}) runs over the $2N$ indices corresponding to $N$ mode pairs with opposite frequencies and identical eigenfunctions. The normalization condition (\ref{orth2}) becomes
\beq
\label{orth4}
\sum_\beta \Big(|a_\beta^{(\alpha)}|^2\omega_\beta^2 + a^{(\alpha)^*}_\beta a^{(\alpha)}_{-\beta} \omega_{-\beta}^2 \Big) + \omega_\alpha^2 \sum_\beta \Big(|a_{\alpha,\beta}|^2 + a^{(\alpha)^*}_\beta a^{(\alpha)}_{-\beta} \Big) = 2 \omega_\alpha^2,
\eeq
where the subscript $-\beta$ refers to the original mode with
eigenfrequency of opposite sign. The sum in equation
(\ref{orth4}) over the $2N$ modes $\beta$ can be written more simply as
a sum over the $N$ modes with $\omega_\beta > 0$:
\beq
\label{orth5}
\sum_{\omega_\beta > 0} \frac{1}{2} \bigg[1+\frac{\omega_\beta^2}{\big( \omega^{(\alpha)} \big)^2}\bigg] \bigg| a^{(\alpha)}_\beta + a^{(\alpha)}_{-\beta} \bigg|^2 = 1.
\eeq
In the limit $\Omega_s \rightarrow 0$, equation (\ref{orth5}) reduces to
$|a_{\alpha,\beta}|^2 \delta_{\alpha,\beta} =1$.

\section{Three Mode Mixing}
\label{threemode}

In this section we solve for the eigenvalues and eigenvectors of a three mode system coupled through the Coriolis force. We consider Mode 1 to be an f-mode, Mode 2 to be an arbitrary mode mixed with the f-mode, and Mode 3 to be an arbitrary mode that does not mix directly with Mode 1 but does mix with Mode 2. The eigensystem describing these three modes is
\beq
\label{3mode1}
\begin{bmatrix} (\bar{\omega_1}-\omega) &  C_{12} & 0 \\ C_{12}^* & (\bar{\omega_2}-\omega) & C_{23} \\ 0 & C_{23}^* & (\bar{\omega_3}-\omega) \end{bmatrix} \left[ \begin{array}{c} b_1 \\ b_2 \\ b_3 \end{array} \right] = 0,
\eeq
with $\bar{\omega}_1 = \omega_1 + C_{11}$ and likewise for modes 2 and 3. This eigensystem is equivalent to 
\beq
\label{3mode2}
\begin{bmatrix} \Delta_{13} - \Delta & 2 C_{12} & 0 \\ 2C_{12}^* & \Delta_{2}-\Delta & 2C_{23} \\ 0 & 2C_{23}^* & -\Delta_{13}-\Delta \end{bmatrix} \left[ \begin{array}{c} b_1 \\ b_2 \\ b_3 \end{array} \right] = 0,
\eeq
where $\Delta_{13}=\bar{\omega}_1-\bar{\omega}_3$, $\Delta_2 = 2\bar{\omega}_2 - \bar{\omega}_1 - \bar{\omega}_3$, and $\Delta = 2\omega - \bar{\omega}_1-\bar{\omega}_3$. 

We now examine the specific case where modes 1 and 3 are degenerate, i.e., $\Delta_{13}=0$. The characteristic equation for this case is
\beq
\Delta\Big[\Delta^2 - \Delta_2\Delta - 4 (|C_{12}|^2 + |C_{23}|^2)\Big] = 0.
\eeq
The eigenvalues are $\Delta_0 = 0$ and $\Delta_\pm = \frac{1}{2} \Big[ \Delta_2 \pm \sqrt{\Delta_2^2 + 16 (|C_{12}|^2 + |C_{23}|^2)}\Big]$. These frequencies correspond to $\omega^{(0)} = \bar{\omega}_1 = \bar{\omega}_3$ and
\beq
\omega^{(\pm)} = \frac{2\bar{\omega}_2+\bar{\omega}_1+\bar{\omega}_3}{4} \pm \frac{1}{4} \bigg[\Delta_2^2 + 16 (|C_{12}|^2 + |C_{23}|^2) \bigg]^{1/2}.
\eeq
The associated (unnormalized) eigenvectors are 
\beq
{\bf b}^{(0)} = \left[ \begin{array}{c} C_{23} \\ 0 \\ -C_{12}^* \end{array} \right] 
\eeq
and
\beq
{\bf b}^{(\pm)} = \left[ \begin{array}{c} 2C_{12} \\ \Delta_\pm \\ 2C_{23}^* \end{array} \right].
\eeq

In all cases, the mode with frequency $\omega^{(0)}$ is a superposition of Modes 1 and 3, with the relative weights determined by the relative values of $|C_{12}|$ and $|C_{23}|$. If $|C_{12}|\ll|C_{23}|$, i.e., Mode 1 is essentially decoupled from the system, then this mode is a slightly perturbed version of Mode 1. If $|C_{12}|\gg|C_{23}|$, i.e., Mode 3 is essentially decoupled from the system, then this mode is a slightly perturbed version of Mode 3.

If the modes are minimally mixed, [$16(|C_{12}| + |C_{23}|) \ll |\Delta_2|$], two modes have frequency $\omega \simeq \bar{\omega}_1=\bar{\omega}_3$ and one has $\omega \simeq \bar{\omega_2}$, as we would expect for non-mixed modes. However, the two modes with $\omega \simeq \bar{\omega}_1=\bar{\omega}_3$ may still be strongly mixed. In the nearly degenerate limit [$16(|C_{12}| + |C_{23}|) \gg |\Delta_2|$], all three modes are mixed with one another, with the weights determined by the relative coupling coefficients. 

It is important to note that if $|C_{12}|$ and $|C_{23}|$, have similar magnitudes, there will always exist at least two modes that are strong superpositions of Modes 1 and 3, regardless of the value of $\Delta_2$. This entails that if a mode (in our example Mode 3) is nearly degenerate with the f-mode, it can mix strongly with it (through another mode, in our example Mode 2) even if it does not mix directly with the f-mode. The mode that serves as intermediary (Mode 2) need not be similar in frequency to Mode 1 and 3. In Section \ref{2and3mode}, we show that systems in which $|C_{12}|$ and $|C_{23}|$ have very different magnitudes can still exhibit strong mixing between Modes 1 and 3. In this case, the strong mixing does not occur when $\bar{\omega}_1 \simeq \bar{\omega}_3$, although it does occur when there are two eigenfrequencies with $\omega \simeq \bar{\omega}_1$.  We should therefore expect that modes very near in frequency to the f-mode will mix strongly with it, regardless of their original characteristics.

\section{Effect of Modes on the Rings}
\label{rings}

The characteristics of density waves launched at Lindblad resonances are characterized in Goldreich \& Tremaine (1979) and in Greenberg \& Brahic (1984). Here we summarize the relevant results and apply them to Saturn's ring system. The wave-like response of the disk near (but not exactly at) the Lindblad resonance is a traveling wave of the form 
\beq
\label{ringpot}
\delta \Phi \simeq -\Psi \sqrt{2\pi |z|} \exp{\Big[ ix^2/(2z)\Big]}.
\eeq
Here, $\delta \Phi$ is the wave-like gravitational perturbation produced in the rings (not to be confused with $\delta \Phi_\alpha$ or $\delta \Phi^{(\alpha)}$ associated with a given oscillation mode), $\Psi$ is the effective perturbing potential (see equation \ref{psi}), $x= (r-r_L)/r_L$ is the fractional distance away from the Lindblad resonance, and
\beq
z = \frac{2 \pi G \Sigma}{r \mathcal{D}},
\eeq
is approximately the square of the radial wavelength. Here, $\Sigma$ is the unperturbed ring surface density, and $\mathcal{D} \simeq 3(m-1) \Omega^2$ near the Lindblad resonance. The value of $z$ is of order $z \sim M_{\rm rings}/M \sim 10^{-9}$, where $M_{\rm rings}$ is the mass of the C-ring. The wavelength is very short compared to the Lindblad radius, thus, the waves are very tightly wound. The wavelength decreases away from the Lindblad radius, although the waves typically damp out after $\sim \! \! 10$ wavelengths. This form of the response is accurate for $c_s/(\Omega r) \ll 1$, $G\Sigma/(\Omega^2 r) \ll 1$, $c_s^2/(G\Sigma r) \ll 1$, where $c_s$ is the typical random velocity of ring particles. These are all excellent approximations for Saturn's rings. 

A fairly accurate location of the resonant location of a mode is obtained by solving equation (\ref{lindblad}), including corrections due to the gravitational moments of Saturn. The solution is 
\beq
\label{rL}
r_L \simeq r_{L0} \bigg[1 - \frac{1+m}{2(1-m)}J_2\left(\frac{R}{r_0}\right)^2 +\frac{5(3+m)}{8(1-m)}J_4\left(\frac{R}{r_0}\right)^4\bigg] + \mathcal{O}\big(J_6,J_2^2\big),
\eeq
where $r_{L0}$ is given by equation (\ref{r0}), and for Saturn $J_2=1.633\times 10^{-2}$ and $J_4=-9.2\times 10^{-4}$ (Guillot 2005). The effective forcing potential is 
\beq
\label{psi}
\Psi = A \bigg[\frac {d}{d \ln r} + \frac{2m \Omega}{m\Omega+\sigma_\alpha}\bigg] \delta \Phi_\alpha(r_L),
\eeq
where $A$ is the mode amplitude. Each component $l'$ of the potential has the form $\delta \Phi_{\alpha,l'} \propto r^{-(l'+1)}$, resulting in $\Psi_{\alpha,l'} = (2m' - l'-1) \delta \Phi_{\alpha,l'}$ (see also equation \ref{respot2}).

The associated density perturbation is 
\begin{align}
\label{denspert}
\delta \Sigma &\simeq \frac{i}{2\pi G r^{1/2}} \frac{d}{dr} \Big(r^{1/2} \delta \Phi \Big)\nonumber \\
&\simeq \bigg|\frac{3(1-m)}{4\pi^2} \frac{\Omega^2 |\Psi|^2}{G^3 \Sigma r_L} \bigg|^{1/2} x \exp{\Big( i \Big[ x^2/(2z) + m \phi + \sigma t \Big] \Big)}. 
\end{align}
This equation applies in the inviscid limit and shows that the density perturbation becomes larger with increasing distance away from resonance. In reality, the waves damp out and (in Saturn's rings) the density perturbations typically decrease after one or two wavelengths away from resonance.\footnote{Equation (\ref{denspert}) applies in the linear regime. Detectable waves in Saturn's rings often have density perturbations of order unity and are relatively non-linear, causing their optical depth variations to be cuspy. We ignore this issue in this work, although it may affect estimates of mode amplitude.} The associated perturbation in optical depth is $\delta \tau = \kappa_m \delta \Sigma$, where $\kappa_m$ is the local mass extinction coefficient (opacity) of the rings, and we have assumed that it is independent of $\Sigma$. 

We would like to use the observed variation in optical depth to estimate the amplitudes of the modes driving the waves. Since the wave amplitude is damped by viscous effects, we must judiciously choose a location $x$ at which to evaluate equation (\ref{denspert}). This location must be far enough from the resonance such that equation (\ref{ringpot}) is applicable, but close enough so that the wave has not been significantly damped. Evaluating equation (\ref{denspert}) at $x=\Delta r/r_L$, we have 
\beq
\label{taupert}
|\delta \tau| \approx \bigg|\frac{3(1-m)}{4\pi^2} \frac{\kappa_m^2 \Omega^2 |\Psi|^2}{G^3 \Sigma r_L} \bigg|^{1/2} \bigg| \frac{\Delta r}{r_L} \bigg|,
\eeq
Thus, the amplitude of the mode is 
\beq
\label{Amp}
|A| \approx \bigg|\frac{3(1-m)}{4\pi^2} \frac{\kappa_m^2 \Omega^2 |\bar{\Psi}|^2}{G^3 \Sigma r_L} \bigg|^{-1/2} \bigg|\frac{r_L}{\Delta r}\bigg| |\delta \tau|,
\eeq
with 
\beq
\label{psibar}
\bar{\Psi} = \sum_{\beta} a_{\beta} W_{\beta} \delta \Phi_{\beta}(R).
\eeq
To estimate mode amplitudes from HN13, we typically evaluate $\delta \tau$ near its maximum, about one wavelength away from the resonant radius, at $\Delta r \sim 5 {\rm km}$. This procedure assumes that damping has had a negligible effect on wave amplitude within the first wavelength. 

Finally, to calculate optical depth fluctuations produced by modes in our Saturn models, we first calculate the amplitude of a mode required to produce the largest observed $m=-3$ wave in Saturn's rings. We use the observed values of $\kappa_m$ and $\sigma_m$ from HN13, and we assume the wave is generated by an unmixed $l=3$, $m=-3$ f-mode with $\delta \Phi_\alpha (R) \sim  1$, determined from our mode calculations. This procedure typically results in a dimensionless amplitude $|A_3|\simeq 10^{-9}$. We then assume energy equipartition amongst the oscillation modes, such that the amplitudes of modes in our model are determined by $\omega_\alpha^2 |A_\alpha|^2 = \omega_3^2 |A_3|^2$. The optical depth variation is then calculated according to equation (\ref{taupert}), and is evaluated at $\Delta r \sim 5 {\rm km}$.

\end{document}